\documentclass[preprint2]{aastex}
\usepackage[utf8]{inputenc}

\usepackage{natbib}
\usepackage{graphicx}
\usepackage{amsmath}
\usepackage{booktabs}
\usepackage{longtable}
\usepackage{xspace}
\usepackage{hyperref}
\usepackage[T1]{fontenc}
\usepackage{lipsum}
\usepackage{comment}
\usepackage{xcolor}

\graphicspath{{Figures/}} 

\newcommand{\ms}{\ensuremath{\rm m\,s^{-1}}}

\newcommand{\gcmc}{\ensuremath{\rm g\,cm^{-3}}}
\newcommand{\gcc}{\gcmc}

\newcommand{\teff}{\ensuremath{T_{\rm eff}}}

\newcommand{\feh}{[Fe/H]}

\newcommand{\rsun}{\ensuremath{R_\sun}}
\newcommand{\msun}{\ensuremath{M_\sun}}

\newcommand{\rstar}{\ensuremath{R_\star}}
\newcommand{\mstar}{\ensuremath{M_\star}}

\newcommand{\rpl}{\ensuremath{R_{\rm p}}}
\newcommand{\mpl}{\ensuremath{M_{\rm p}}}

\newcommand{\rhopl}{\ensuremath{\rho_{\rm p}}}

\newcommand{\rjup}{\ensuremath{R_{\rm J}}}
\newcommand{\mjup}{\ensuremath{M_{\rm J}}}

\newcommand{\rearth}{\ensuremath{R_\earth}}
\newcommand{\mearth}{\ensuremath{M_\earth}}

\newcommand{\msini}{\ensuremath{M \sin{i}}}
\newcommand{\mplsini}{\ensuremath{\mpl\sin{i}}}

\newcommand{\Kepler}{\textit{Kepler}}

\newcommand{\ecosw}{\ensuremath{e\rm{cos}\omega}}
\newcommand{\esinw}{\ensuremath{e\rm{sin}\omega}}
\newcommand{\secw}{\ensuremath{\sqrt{e}\rm{cos}\omega}}
\newcommand{\sesw}{\ensuremath{\sqrt{e}\rm{sin}\omega}}
\newcommand{\K}{\ensuremath{K}}
\newcommand{\shk}{\ensuremath{S_{\rm{HK}}}}

\newcommand{\pomega}{\varpi}
\newcommand{\Pbval}{\ensuremath{10.91647\pm0.00014\,\rm{days}}}
\newcommand{\Pcval}{\ensuremath{22.2649\pm0.0007 \,\rm{days}}}
\newcommand{\Pdval}{\ensuremath{1403\pm14\,\rm{days}}}

\newcommand{\Rbval}{\ensuremath{3.438\pm0.075\,\rearth}}

\newcommand{\Mbval}{\ensuremath{9.5\pm1.2\,\mearth}}
\newcommand{\Mcval}{\ensuremath{214.0\pm5.3\,\mearth}}
\newcommand{\mdval}{\ensuremath{965\pm44\,\mearth}}
\newcommand{\mdvaljup}{\ensuremath{3.04\pm0.13\,\mjup}}

\newcommand{\PcRV}{\ensuremath{22.2695\pm0.0045}\,\rm{days}}
\newcommand{\McRV}{\ensuremath{208\pm12}\,\mearth}

\newcommand{\MdRV}{\ensuremath{1000\pm48\,\mearth}}

\newcommand{\mstarval}{\ensuremath{0.985^{+0.027}_{-0.022}\,\msun}}
\newcommand{\rstarval}{\ensuremath{0.900\pm0.022\,\rsun}}
\newcommand{\newmstarval}{\ensuremath{0.990\pm0.024\,\msun}}
\newcommand{\newrstarval}{\ensuremath{0.897\pm0.016\,\rsun}}

\newcommand{\ec}{\ensuremath{0.0572\pm0.0004}}
\newcommand{\edval}{\ensuremath{0.41\pm0.03}}

\newcommand{\nrvs}{\ensuremath{44}}
\newcommand{\minsnr}{\ensuremath{50}} 

\newcommand{\Pcavgshort}{22.26 days}

\newcommand{\fehval}{\ensuremath{0.27\pm0.06}}

\begin{document}
\title{The Discovery of the Long-Period, Eccentric Planet Kepler-88 d and System Characterization with Radial Velocities and Photodynamical Analysis}
\author[0000-0002-3725-3058]{Lauren M. Weiss}
\affiliation{Institute for Astronomy, University of Hawai`i, 2680 Woodlawn Drive, Honolulu, HI 96822, USA}
\author[0000-0003-3750-0183]{Daniel~C.~Fabrycky}
\affiliation{Department of Astronomy and Astrophysics, University of
Chicago, 5640 S. Ellis Ave, Chicago, IL 60637, USA}
\author[0000-0002-0802-9145]{Eric Agol}
\affil{Department~of~Astronomy, University~of~Washington, Seattle, WA}
\author[0000-0002-4535-6241]{Sean M. Mills}
\affiliation{California Institute of Technology, 1200 E California Blvd, Pasadena, CA 91125, USA}
\author[0000-0001-8638-0320]{Andrew W. Howard}
\affiliation{California Institute of Technology, Pasadena, CA 91125, USA}
\author[0000-0002-0531-1073]{Howard Isaacson}
\affiliation{501 Campbell Hall, University of California at Berkeley, Berkeley, CA 94720, USA}
\author[0000-0003-0967-2893]{Erik A Petigura}
\affiliation{Department of Physics \& Astronomy, University of California Los Angeles, Los Angeles, CA 90095, USA}
\author[0000-0003-3504-5316]{Benjamin Fulton}
\affiliation{NASA Exoplanet Science Institute, MC 314-6, 1200 E California Blvd, Pasadena, CA 91125, USA}
\author[0000-0001-8058-7443]{Lea Hirsch}
\affiliation{Varian Physics, Room 108, 382 Via Pueblo Mall, Stanford, CA 94305, USA}
\author[0000-0002-5658-0601]{Evan Sinukoff}
\affiliation{Institute for Astronomy, 2680 Woodlawn Dr., Honolulu, HI 96822, USA}

\begin{abstract}
We present the discovery of Kepler-88 d ($P_d =$ \Pdval, $\msini_d =$ \mdval = \mdvaljup, $e_d = \edval$) based on six years of radial velocity (RV) follow-up from the W. M. Keck Observatory HIRES spectrograph.  Kepler-88 has two previously identified planets.  Kepler-88 b (KOI-142.01) transits in the NASA \Kepler\ photometry and has very large transit timing variations.  \citet{Nesvorny2013} perfomed a dynamical analysis of the TTVs to uniquely identify the orbital period and mass of the perturbing planet (Kepler-88 c), which was later was confirmed with RVs from the Observatoire de  Haute-Provence \citep[OHP,][]{Barros2014}.  To fully explore the architecture of this system, we performed photodynamical modeling on the \Kepler\ photometry combined with the RVs from Keck and OHP and stellar parameters from spectroscopy and Gaia.  Planet d is not detectable in the photometry, and long-baseline RVs are needed to ascertain its presence.  A photodynamical model simultaneously optimized to fit the RVs and \Kepler\ photometry yields the most precise planet masses and orbital properties yet for b and c: $P_b = \Pbval$, $M_b=\Mbval$, $P_c=\Pcval$, and  $M_c=\Mcval$.  The photodynamical solution also finds that planets b and c have low eccentricites and low mutual inclination, are apsidally anti-aligned, and have conjunctions on the same hemisphere of the star.  Continued RV follow-up of systems with small planets will improve our understanding of the link between inner planetary system architectures and giant planets. 

\end{abstract}


\section{Introduction}
The NASA \Kepler\ Mission detected hundreds of systems with multiple transiting planets \citep{Lissauer2011_multis, Fabrycky2014,Rowe2014, Lissauer2014b}, providing insight into one of the most common modes of planet formation.  One unexpected attribute of the \Kepler\ planetary systems is that planets in or very near mean-motion resonances are rare \citep{Fabrycky2014}.  The prevalence of planets that are not in mean-motion resonances seems at odds with examples from our solar system (e.g., the Galilean moons) and resonant chains of giant exoplanets detected in radial velocity surveys \citep[e.g.,][]{Marcy2001}.  Resonant architectures are expected to arise when planet pairs migrating convergently become trapped in the energetically favorable configuration of mean motion resonance.  Since viscous migration in a disk is often invoked to explain the prevalence of volatile-containing planets within 1 AU, the dearth of resonant planetary architectures in the compact Kepler planetary systems is an unsolved puzzle.

\begin{figure}
    \centering
    \includegraphics[width=0.5\textwidth]{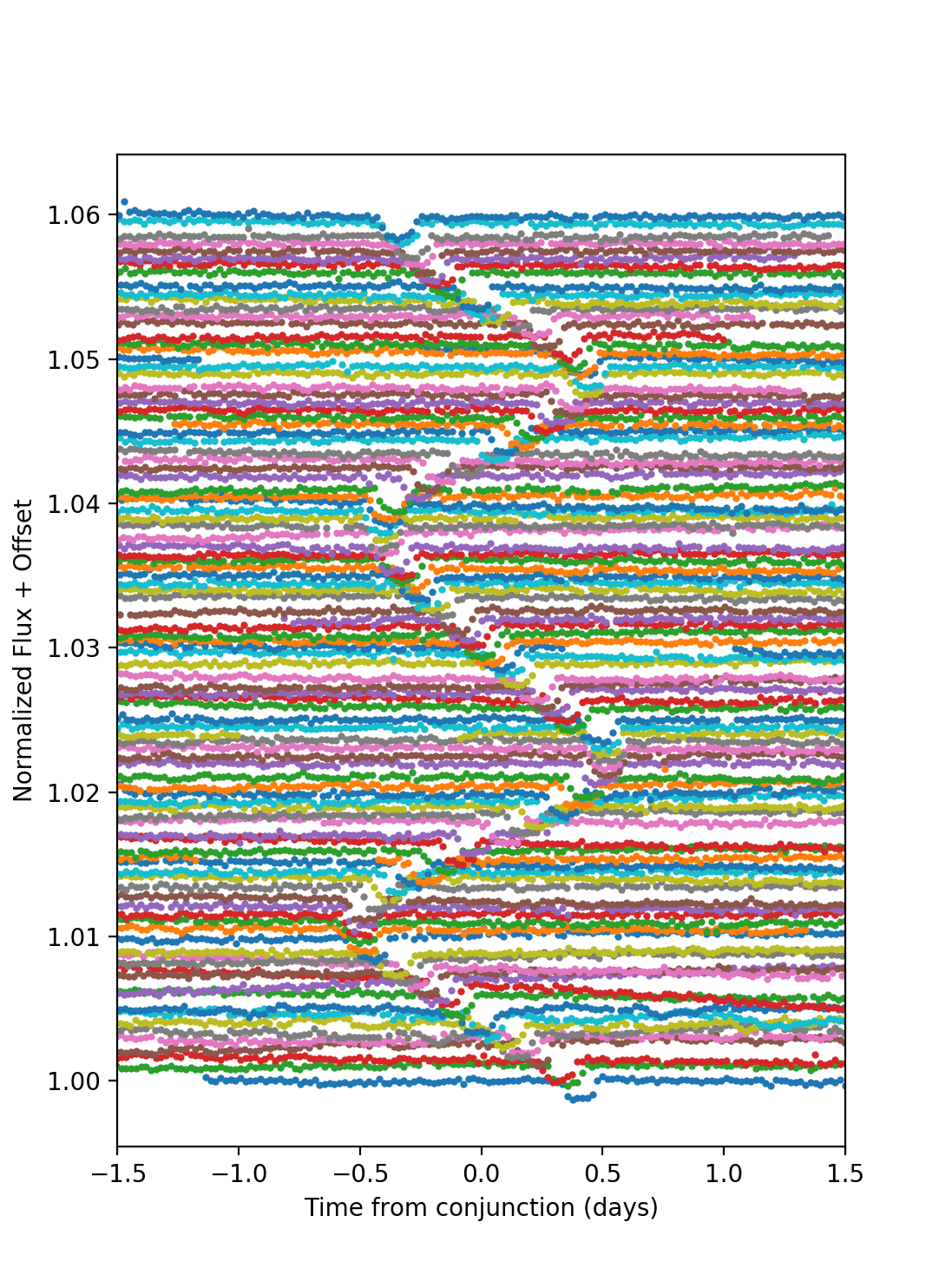}
    \caption{The Kepler long-cadence photometry of Kepler-88 near expected times of conjunction for Kepler-88 b ($P=10.95$ days), with individual transits offset vertically (epoch increases from bottom to top, and the colors disambiguate adjacent epochs).  The TTVs of amplitude 0.5 days are readily identifiable.}
    \label{fig:riverplot}
\end{figure}

Kepler-88 (KOI-142) is a rare example of a planetary system very near a mean-motion resonance.  The system has only one transiting planet, Kepler-88 b (KOI-142.01), a sub-Neptune-sized planet with orbital period of 10.95 days. Kepler-88 b is perturbed by a non-transiting giant planet with a period of 22.26 days, Kepler-88 c \citep[KOI-142.02,][]{Nesvorny2013}.  The resonant conjunctions of the sub-Neptune and giant planet produce large transit timing variations (TTVs), which have an amplitude of half a day (5\% of the orbital period of the transiting planet, see Figure \ref{fig:riverplot}).  These very large TTVs led to the nickname ``The King of TTVs'' for the Kepler-88 system \citep{Steffen2012} and have been identified in various TTV catalogs \citep[e.g.,][]{Ford2011,Steffen2012,Mazeh2013,Holczer2016}.  

The Kepler-88 b TTVs were first explained dynamically in \citet[][N13]{Nesvorny2013}. In an N-body dynamical fit, they found that (1) the perturber of the Neptune-sized planet is at $22.3397^{+0.0021}_{-0.0018}$ days, (2), the mass of the perturber is $198.8^{+9.2}_{-10.6}\,\mearth$, (3) the eccentricities of the 11-day and 22-day planet are small but non-zero ($e_b=0.05596^{+0.00048}_{-0.00034}$, $e_c=0.0567^{+0.0010}_{-0.0013}$), and (4) the orbits of the resonant planets are apsidally anti-aligned ($\Delta \pomega=180\pm2^{\circ}$).  N13 also found non-negligible transit duration variations (TDVs) of the transiting planet, which provided a constraint on the mutual inclination of the two planets.

Shortly thereafter, \citet{Barros2014} used the Observatoire de Haute-Provence (OHP) telescope and SOPHIE high-resolution echelle spectrograph to measure radial velocities (RVs) of the Kepler-88 system.  With one season of RVs, they confirmed the presence of a $241^{+102}_{-51}$ \mearth\ planet with an orbital period of $22.10\pm0.25$ days.  This was the first time that RVs confirmed an accurate and precise prediction of the location and mass of a non-transiting exoplanet from TTVs.

In this paper, we present RVs of Kepler-88 from the W. M. Keck Observatory HIRES spectrograph taken between the years 2013 and 2020.  Our RVs confirm the existence, mass, and orbital period of the giant planet at \Pcavgshort.  We also detect another giant planet in the system, Kepler-88 d, at an orbital period of \Pdval, with minimum mass ($\msini_d$) of \mdval\ and an eccentricity of \edval.  The high mass and eccentricity of Kepler-88 d indicate that it has likely been an important dynamical component in this planetary system's history.  To identify accurate dynamical parameters for all of the known bodies in the system, we simultaneously fit the Kepler photometry, Keck-HIRES RVs, and OHP-SOPHIE RVs of Kepler-88 with multiple N-body codes.

This paper is organized as follows:  In \S2, we present our observing strategy and the Keck-HIRES RVs, literature RVs, and stellar properties.  In the following sections, we explore the RV data with increasingly complex models and supplementary data.  In \S3, we present a three-planet Keplerian model to the RVs.  In \S4, we present the results of a simultaneous N-body fit to the RVs and TTVs.  In \S5, we perform simultaneously an N-body fit to the RVs and \Kepler\ photometry (a photodynamical fit).  In \S6, we present the main results from our analyses.  In \S7, we discuss how our results affect our interpretation of the history of this planetary system, and how this system adds to the small but growing list of systems with characterizations from both RV and TTV analyses.  In \S8 we conclude.

\section{Keck-HIRES Spectra}
\subsection{Radial Velocities}
We obtained \nrvs\ RVs of Kepler-88 on the HIRES spectrograph \citep{Vogt1994} at the W. M. Keck Observatory between the years 2013 and 2020.  We used the standard HIRES setup of the California Planet Search \citep[see][for details]{Howard2010_CPS}.  Spectra were obtained using HIRES in the red-collimation mode with a warm molecular iodine gas cell in the light path for wavelength calibration.  We used the C2 decker ($0.\arcsec86 \times 14\arcsec$, R=60,000) to enable sky-subtraction for this relatively faint ($V=13.8$) target.  Since the target was faint, we only observed in good conditions (seeing $< 1.\arcsec5$, clear to thin clouds).  For each spectrum, we achieved a signal to noise ratio of at least \minsnr\ to ensure that our Doppler pipeline would deliver RVs with errors of $< 10\,\ms$ \citep{Howard2016}.

We observed an iodine-free template spectrum bracketed by observations of rapidly rotating B-type stars to enable a deconvolution of the stellar spectrum from the spectrograph PSF. We then forward-modeled our RV spectra with the deconvoled template stellar spectrum plus a night-specific model of our PSF convolved with an atlas iodine spectrum.  We also used the blue HIRES chip to extract a Mt. Wilson \shk\ value for each HIRES observation.  Our Keck-HIRES RVs and \shk\ values, plus the SOPHIE RVs from the literature, are presented in Table \ref{tab:rvs}.
\begin{deluxetable}{lcccc}




\tablecaption{Kepler-88 RVs\label{tab:rvs}
}

\tablehead{\colhead{Time} & \colhead{RV} & \colhead{$\sigma_\mathrm{RV}$} & \colhead{$S_{\rm{HK}}$} & \colhead{Inst} \\ 
\colhead{(BJD - 2454900)} & \colhead{(m/s)} & \colhead{(m/s)} & \colhead{}& \colhead{} } 

\startdata
1575.047908 & -35.7 & 2.8 & 0.142 & HIRES \\
1575.40947 & 42.0 & 10.0 &  & SOPHIE \\
1575.992695 & -36.7 & 2.4 & 0.155 & HIRES \\
1577.892731 & -52.9 & 2.5 & 0.147 & HIRES \\
1579.027079 & -81.7 & 2.6 & 0.122 & HIRES \\
...         & ... & ... & ... & ... \\
\enddata



\tablecomments{Times are in BJD - 2454900.0.  SOPHIE RVs are from \citet{Barros2014}; HIRES RVs are from this work.  The SOPHIE RVs have had 20465.0 \ms added, with respect to the values published in \citet{Barros2014}, for easier zero-point calibration.  The RV uncertainties do not include RV jitter.  The full table is available in machine readable form.  The first few lines are shown here for content and format.}

\end{deluxetable}

\subsection{Stellar Parameters}
The stellar properties of Kepler-88 were determined based on our high signal-to-noise template spectrum in combination with the \textit{Gaia} parallax and 2MASS photometry \citep{FultonPetigura2018}.  The stellar temperature is $5466\pm60$\,K and with \feh = \fehval, the star is slightly metal-rich.  The star has a similar mass but slightly smaller radius than the sun (\mstar = \mstarval, \rstar = \rstarval).  

Because the transit of planet b combined with dynamical information about the planet constrains the density of the star, we used a photodynmical fit to update the stellar characterization \citep[e.g.,][see section \ref{sec:photodynamical}]{Vanderburg2017}.  We used the best-fit values and uncertainties for the stellar mass and radius from \citet{FultonPetigura2018} as priors in our photodynamical fit.  After our photodynamical fit, the best-fit stellar mass and radius are \mstar = \newmstarval\ and \rstar = \newrstarval.  The precision of the stellar radius determination was improved through the photodynamical fit, suggesting that the transits provide information about the stellar density and hence the stellar radius.\footnote{The stellar mass was essentially unchanged, which is the expected behavior from \texttt{Phodymm} \citep[]{Mills2016}.}

Of the stellar parameters reported here, only the stellar mass is dependent on isochrone fitting \citep[see][for details]{FultonPetigura2018}.  We caution that the formal error in the mass reported here does not account for systematic differences between the stellar isochrones formulated by different research groups, and so the reported error in the stellar mass (and hence density) might be underestimated.

\section{Keplerian fit \label{sec:keplerian}}
\begin{figure}
    \centering
    \includegraphics[width=0.5\textwidth]{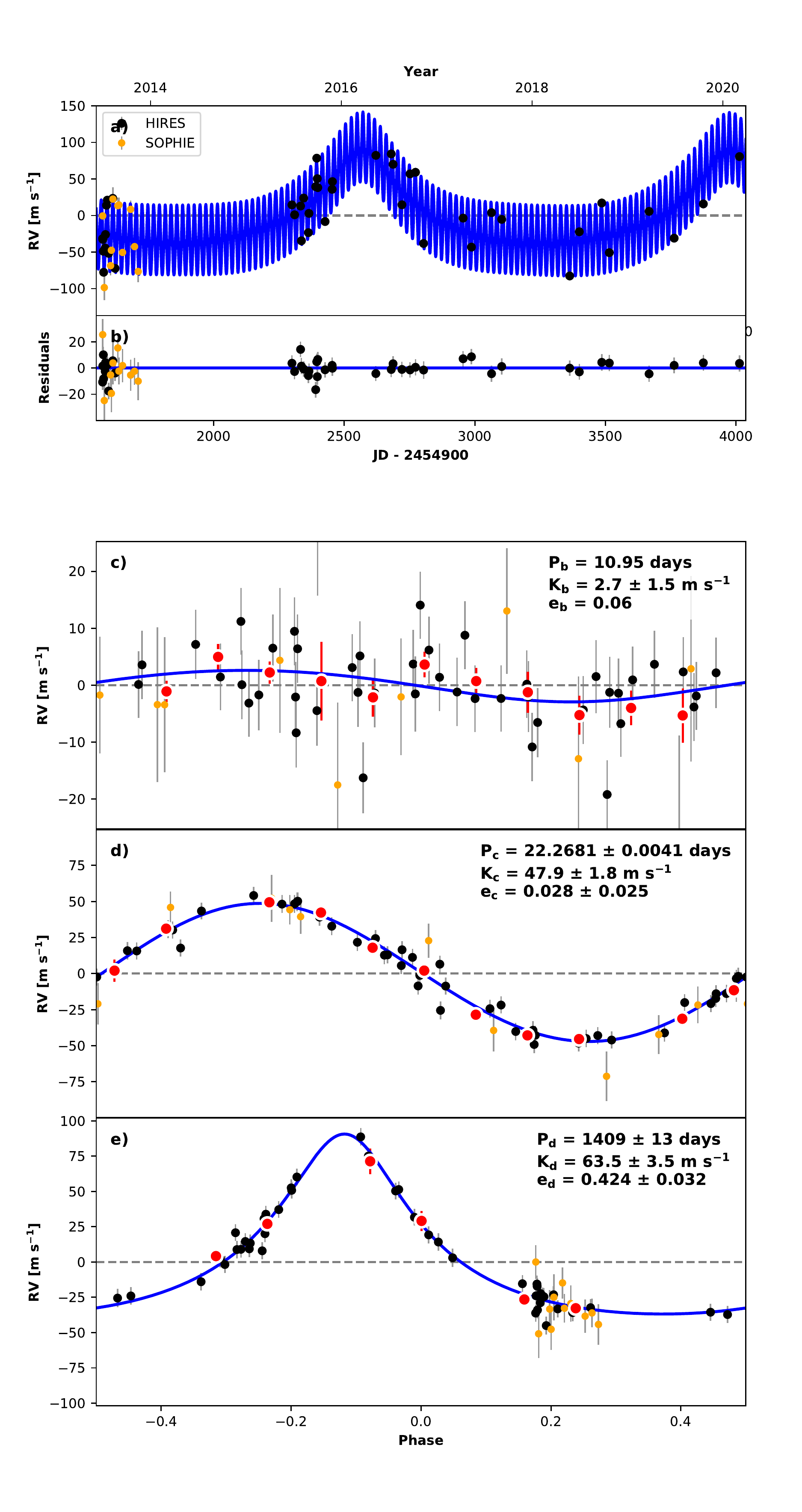}
    \caption{a: The RVs of Kepler-88 from Keck-HIRES (black) and OHP-SOPHIE (orange), and their errors (including jitter) as a function of time.  The best fit 3-planet Keplerian solution is shown in blue.  b: The residuals.  c: The RVs phase-folded to the linear ephemeris of planet b, with the other Keplerian signals removed.  Binned average RVs and their uncertainties are shown in red.  The RVs alone do not detect the mass of planet b, but planet b clearly exists from its transits.  d: Same as above, but for planet c, which does not transit.  e: Same as above, but for planet d, which does not transit.  With a period of 1400 days, planet d is not detected in the TTVs, and so only the RVs provide a useful determination of its orbital properties.}
    \label{fig:koi142-radvel}
\end{figure}

The RVs of Kepler-88 show long-term variation from a planetary companion at $\sim 4$ years (see Figures \ref{fig:koi142-radvel}, \ref{fig:koi142-2pl}, and \ref{fig:koi142-3pl}).  The discovery of this companion is the result of the long baseline (currently six years) of Keck-HIRES RVs.  The 4-year RV variation does not correlate with \shk\ variability, strongly disfavoring a stellar activity cycle as the source of the RV signal.

To obtain initial estimates of the orbital properties of all three planets, we fit the RVs from both HIRES and SOPHIE with a 3-planet Keplerian model.  Since the innermost planet is very low mass, we fixed its orbital period and transit time at the best linear-ephemeris values as determined from the \citet{Holczer2016} TTVs ($P_b = 10.95$ days), and kept its eccentricity fixed.  We allowed the five orbital elements $P$, $T_{\rm{p}}$, \secw, \sesw, and \K\ to vary for planets c and d, as well the HIRES RV zeropoint ($\gamma_{\mathrm{HIRES}}$), the SOPHIE RV zeropoint ($\gamma_{\mathrm{SOPHIE}}$), and the RV jitter for each telescope ($\sigma_{\mathrm{HIRES}}$), ($\sigma_{\mathrm{SOPHIE}}$).  Our priors were $0<e<1$ and $K>0$ for all planets.  We explored these parameters with a Markov-Chain Monte Carlo (MCMC) analysis, the results of which are in Tables \ref{tab:radvel-params} and \ref{tab:radvel-derived}.

The RVs place tight constraints on planet masses \msini$_c$ (\McRV) and \msini$_d$ (\MdRV), but provide very little information about $M_b$.  This is because the transiting planet is small and the star is faint; many RVs are needed in this regime to obtain accurate and precise planet masses.  As we show below, however, incorporating the TTVs or a full photodynamical model dramatically improves our constraint on the masses and orbits of Kepler-88 b and c with respect to the RV-only solution.

\begin{deluxetable*}{lrrr}
\tablecaption{RV Only Keplerian MCMC Posteriors }
\tablehead{
  \colhead{Parameter} & 
  \colhead{Credible Interval} & 
  \colhead{Maximum Likelihood} & 
  \colhead{Units}
}
\startdata
\sidehead{\bf{Orbital Parameters}}
  $P_{b}$ & $\equiv10.9531$ & $\equiv10.9531$ & days \\
  $T\rm{conj}_{b}$ & $\equiv175.1591$ & $\equiv175.1591$ & JD \\
  $e_{b}$ & $\equiv0.06$ & $\equiv0.06$ &  \\
  $\omega_{b}$ & $\equiv-3.1306$ & $\equiv-3.1306$ & radians \\
  $K_{b}$ & $2.7^{+1.6}_{-1.4}$ & $2.8$ & m s$^{-1}$ \\
  $P_{c}$ & $22.2681^{+0.0042}_{-0.004}$ & $22.2679$ & days \\
  $T\rm{conj}_{c}$ & $172.28^{+0.46}_{-0.49}$ & $172.27$ & JD \\
  $e_{c}$ & $0.03^{+0.03}_{-0.02}$ & $0.02$ &  \\
  $\omega_{c}$ & $-0.3^{+1.8}_{-1.4}$ & $-0.5$ & radians \\
  $K_{c}$ & $47.9^{+1.9}_{-1.8}$ & $47.9$ & m s$^{-1}$ \\
  $P_{d}$ & $1409^{+14}_{-13}$ & $1409$ & days \\
  $T\rm{conj}_{d}$ & $1325\pm 21$ & $1327$ & JD \\
  $e_{d}$ & $0.424^{+0.031}_{-0.032}$ & $0.422$ &  \\
  $\omega_{d}$ & $0.04^{+0.08}_{-0.075}$ & $0.033$ & radians \\
  $K_{d}$ & $63.5^{+3.5}_{-3.4}$ & $63.7$ & m s$^{-1}$ \\
\hline
\sidehead{\bf{Other Parameters}}
  $\gamma_{\rm SOPHIE}$ & $42.2^{+4.5}_{-4.7}$ & $42.6$ & m s$-1$ \\
  $\gamma_{\rm HIRES}$ & $-4.0\pm 1.3$ & $-4.0$ & m s$-1$ \\
  $\dot{\gamma}$ & $\equiv0.0$ & $\equiv0.0$ & m s$^{-1}$ d$^{-1}$ \\
  $\ddot{\gamma}$ & $\equiv0.0$ & $\equiv0.0$ & m s$^{-1}$ d$^{-2}$ \\
  $\sigma_{\rm SOPHIE}$ & $8.5^{+5.5}_{-5.0}$ & $6.4$ & $\rm m\ s^{-1}$ \\
  $\sigma_{\rm HIRES}$ & $6.57^{+1.0}_{-0.86}$ & $5.41$ & $\rm m\ s^{-1}$ \\
 \enddata
\tablenotetext{}{
  BJD$_0= 2454900$.  
}
\label{tab:radvel-params}
\end{deluxetable*}

\begin{deluxetable*}{lrrr}
\tablecaption{RV-Only Keplerian Derived Posteriors \label{tab:radvel-derived}
}
\tablehead{
  \colhead{Parameter} & 
  \colhead{Credible Interval} & 
  \colhead{Maximum Likelihood} & 
  \colhead{Units}
}
\startdata
 $a_b$ & $0.09604^{+0.00063}_{-0.00066}$ & $0.09687$ &  AU \\
$M_b\sin i$ & $9.3^{+5.5}_{-5.0}$ & $3.4$ & M$_{\oplus}$ \\
$\rho_b$ & $0.9^{+0.7}_{-0.5}$ & $0.4$ &  g cm$^{-3}$ \\
  $a_c$ & $0.154\pm0.001$ & $0.156$ &  AU \\
  $M_c\sin i$ & $0.656^{+0.027}_{-0.026}$ & $0.671$ & M$_{\rm Jup}$ \\
  $a_d$ & $2.45\pm 0.02$ & $2.458$ &  AU \\
  $M_d\sin i$ & $3.15\pm 0.15$ & $3.14$ & M$_{\rm Jup}$ \\
\enddata
\end{deluxetable*}

\section{N-body Fit to TTVs + RVs}
\begin{figure*}[ht!]
\centering
\includegraphics[width=0.75\textwidth]{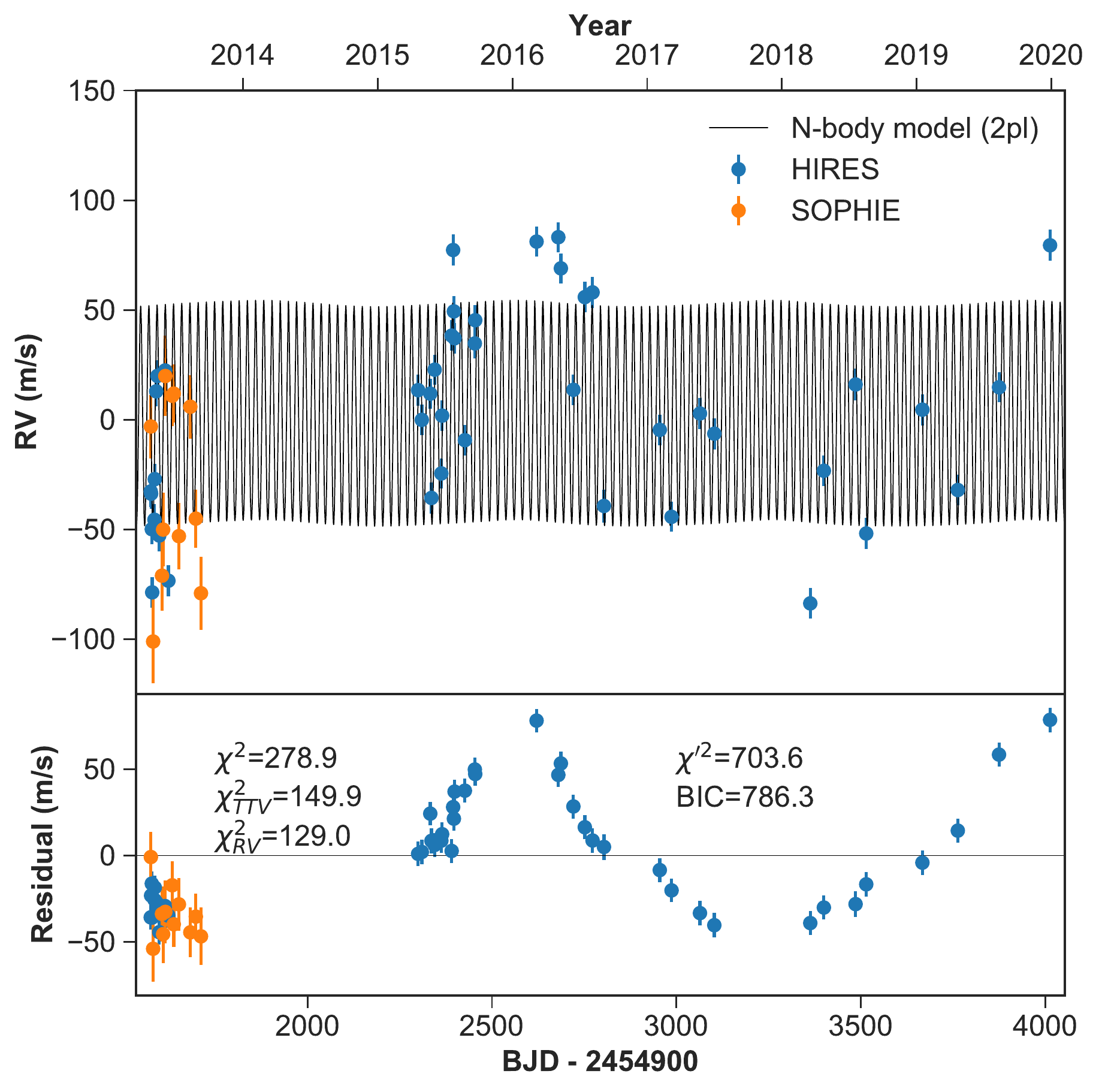}
\caption{Top: Kepler-88 Radial velocities from Keck-HIRES (blue) and OHP-SOPHIE (orange).  The best two-planet N-body fit to the RVS and TTVs (Kepler-88 b and c) is shown in black. Bottom: the RV residuals.  There is a strong residual RV signal near 1400 days.  The $\chi^2$ values of the fit to the TTVs and RVs are given, as are the penalty-adjusted $\chi^{\prime 2}$ and BIC.}
\label{fig:koi142-2pl}
\end{figure*}

\begin{figure*}[ht!]
\centering
\includegraphics[width=0.75\textwidth]{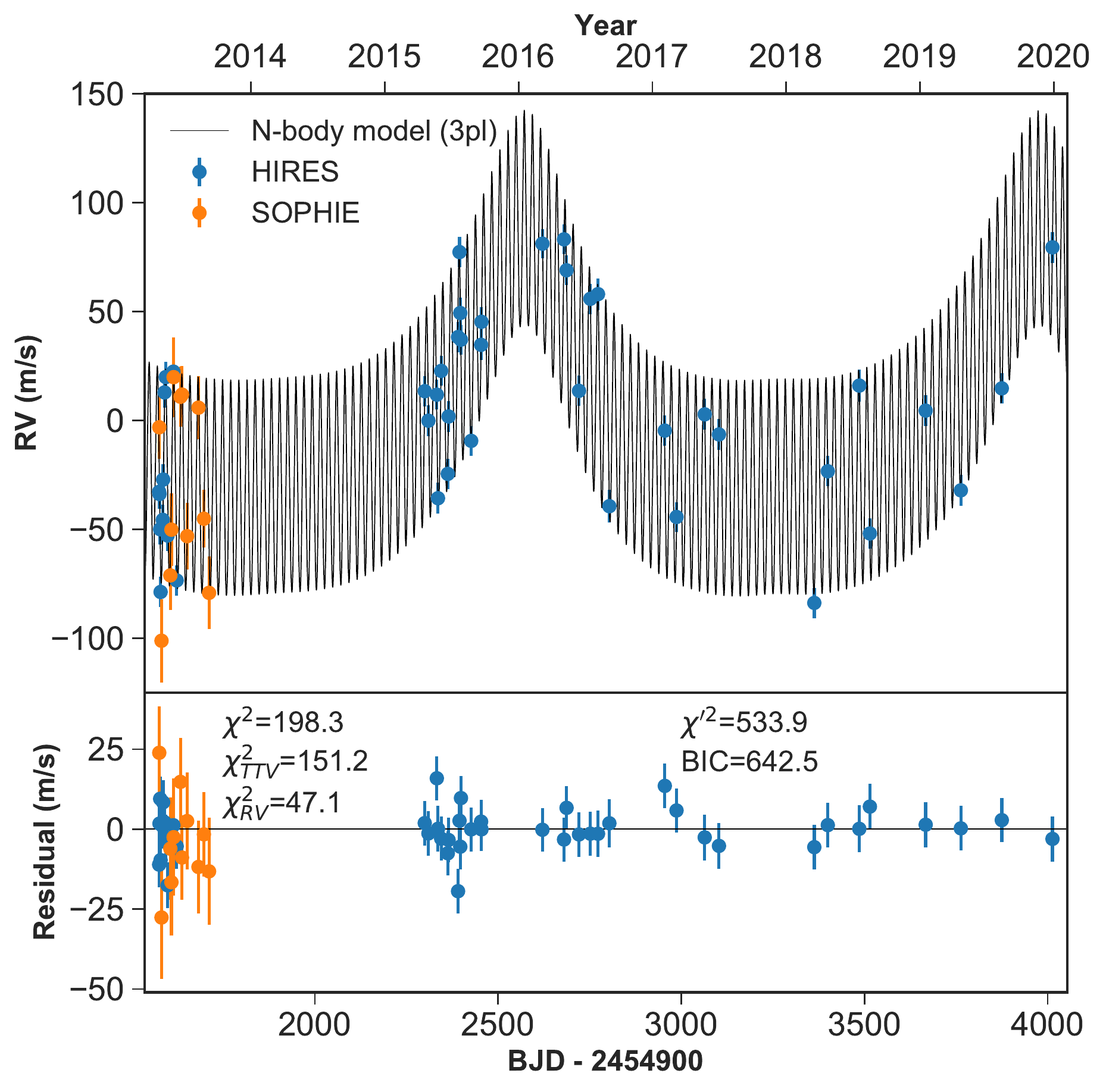}
\caption{Same as Figure \ref{fig:koi142-2pl}, but showing the best three-planet N-body model and residuals (note the different y-axis ranges).  The inclusion of the third planet substantially reduces the RV residuals and improves (reduces) the BIC.}
\label{fig:koi142-3pl}
\end{figure*}

There is an important distinction between an N-body fit (e.g., in Figures \ref{fig:koi142-2pl} and \ref{fig:koi142-3pl}) vs. a multiple-Keplerian fit (e.g., Figure \ref{fig:koi142-radvel}):  N-body fits include planet-planet interactions, whereas Keplerian fits do not.  A detailed N-body analysis is necessary to accurately model the positions and velocities of the Kepler-88 bodies because the two inner planets are near resonance.  

We used the N-body code \texttt{TTVFast} \citep{Deck2014} to simultaneously reproduce the TTVs and RVs in the Kepler-88 system.  For this analysis, we used the TTVs published in \citet{Holczer2016}, which are measured from the \Kepler\ long-cadence photometry.  Our optimization algorithms included least-squares minimization and MCMC analysis.  We considered a two-planet model (planets b and c only) and a three-planet model (planets b, c, and d), fitting the TTVs alone, and then the RVs and TTVs simultaneously. We varied the masses, orbital periods, eccentricities and arguments of pericenter (via parameters \secw\ and \sesw), and mean anomalies for each of the planets at epoch BJD=2454954.62702, as well as the inclination and longitude of ascending node for planet c, and an RV zeropoint jitter for each spectrograph.  We penalized high values of RV jitter in our minimization function $\chi^{\prime 2}$:
\begin{equation}
    \chi^{\prime 2} =  \chi_{\rm{RV}}^2 + \chi_{\rm{TTV}}^2 + \sum_i 2 \rm{ln} \sqrt{2\pi{\sigma^{\prime 2}_i}},
\end{equation}
where $\sigma^\prime_i$ is the quadrature sum of the $i$th individual RV error and the RV jitter of the corresponding spectrograph, and $\chi^2$ is the usual statistic 
\begin{equation}
    \chi^2 = \sum_i \frac{(x_{\mathrm{meas},i} - x_{\mathrm{mod},i})^2}{\sigma_i^{\prime 2}}.
\end{equation}

We compared the goodness of fit of our four models using the Bayesian Information Criterion (BIC):
\begin{equation}
    \mathrm{BIC} = \chi^{\prime 2} + \mathrm{ln}(N) N_{\mathrm{varys}}
\end{equation}
where $N$ is the number of data points (TTV alone or RV + TTV, depending on which data were used) and $N_{\mathrm{varys}}$ is the number of variables.  Note that we use $\chi^{\prime 2}$ instead of $\chi^2$ in calculating the BIC so that the penalty for large RV jitters is included in our model comparison.
The $\chi^2$ values, degrees of freedom, and BICs from our four-way analysis are summarized in Table \ref{tab:bic_compare}.  If only the TTVs are fit, a two-planet model is adequate for fitting the data, based on the similar values of the BIC for the two- and three-planet models ($\Delta$BIC$=26$, in favor of the 2-planet model).  However, in fitting the TTVs combined with the RVs, a three-planet model is strongly preferred, with $\Delta$BIC$=-120$.  To illustrate the better performance of the three-planet model, the RVs are shown with our best two-planet fit (Figure \ref{fig:koi142-2pl}) and our best three-planet fit (Figure \ref{fig:koi142-3pl}).  The TTVs and our best three-planet fit are shown in Figure \ref{fig:koi142-ttv} (upper panel).  The best-fit planet masses and orbits from the TTV and TTV+RV analyses are within $1\sigma$ of the values we find in our photodynamical analysis, which is presented in \S5.  

\begin{deluxetable*}{ccrrrrrrrr}
\tablecaption{Model Comparison\label{tab:bic_compare}}
\tablehead{
  \colhead{Model} & 
  \colhead{Data} & 
  \colhead{N$_{\mathrm{varys}}$}&
  \colhead{$\chi^2_{\rm{TTV}}$} &
  \colhead{$\chi^2_{\rm{RV}}$} &
  \colhead{$\sigma_{\mathrm{HIRES}}$} &
  \colhead{$\sigma_{\mathrm{SOPHIE}}$} &
   \colhead{$\chi^{\prime2}$} &
  \colhead{DOF} & 
  \colhead{BIC}\\
   \colhead{} & 
  \colhead{} & 
  \colhead{}&
  \colhead{} &
  \colhead{} &
  \colhead{(\ms)} &
  \colhead{(\ms)} &
   \colhead{} &
  \colhead{} & 
  \colhead{}}

\startdata
2 planets &	TTVs & 16 & 149.5 &	0 & -&- & 150 & 105	& 226\\
3 planets &	TTVs & 21 & 149.6 & 0 &- &- & 150 & 100	&  250\\
\hline
2 planets &	TTVs + RVs & 16 & 150 & 129  & 20.0 & 9.6 & 703 & 160 & 786 \\
3 planets & TTVs + RVs & 21 & 150 & 50	& 6.6 & 8.1 & 534 & 155 & 642\\
\hline
3 planets, control &      Phot + RVs & 26 & - & 73.0 & 6.5 & 6.2 & 1564297 & 1564457 & 1564668 \\ 
3 planets, $i_c$ flipped &  Phot + RVs & 26 & - & 45.5 & 6.5 & 6.2 & 1564330 & 1564457 & 1564701 \\ 
3 planets, $i_d=30^\circ$ & Phot + RVs & 26 & - & 75.6 & 6.5 & 6.2 & 1564303 & 1564457 & 1564674 \\ 
\enddata

\tablecomments{There are 121 TTV data, 55 RV data, and 1564429 photometric data.  $\sigma_{\mathrm{HIRES}}$ and $\sigma_{\mathrm{SOPHIE}}$ are the HIRES and SOPHIE jitter terms, respectively, which are added to the intrinsic RV errors in quadrature.}
\end{deluxetable*}

\subsection{Chopping Signal}
 In systems that are not very close to resonance and/or have large TTV uncertainties compared to the timing precision, only the low-frequency TTV super-period is detected.  This low-frequency signal contains information about the mass ratio of the planets, but the absolute masses are degenerate with the eccentricities of the planets \citep{Lithwick2012}.  However, in systems with high signal-to-noise TTV measurements like Kepler-88, it is possible to detect a higher-frequency signal: the synodic chopping signal.  This signal abruptly changes direction after conjunctions between the transiting and perturbing planet \citep{Agol&Fabrycky2018}.  The chopping signal is therefore expected to occur at the synodic period, or
\begin{equation}
P_{\rm{chop}} = (1/P_1 - 1/P_2)^{-1}
\end{equation}

\begin{figure*}[h!]
\centering
\includegraphics[width=\textwidth]{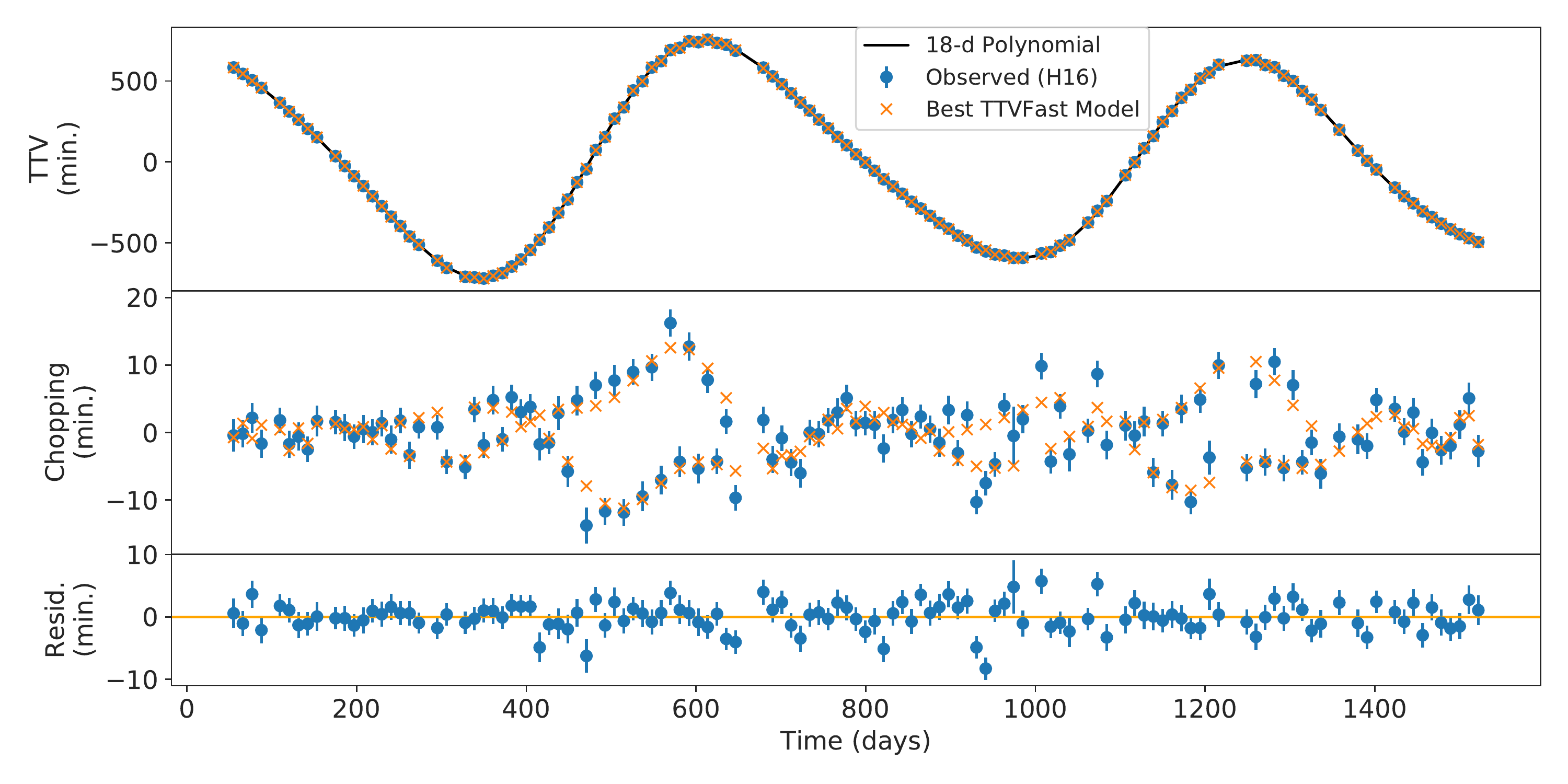}
\caption{Top: The observed TTV signal from \citet{Holczer2016} (blue dots), our best-fit model produced with \texttt{TTVFast} (orange x's), and an 18-degree polynomial fit to the observed TTVs (black line).  Middle: The high-frequency chopping signal was separated from the low-frequency  TTV signal by subtracting the polynomial fit, for both the observed transit times (blue dots) and modeled TTVs (orange x's).  Both the observed and modeled chopping signals have amplitudes that vary with the phase of the TTV super-period.  Bottom: The residuals (observed minus modeled transit time).}
\label{fig:koi142-ttv}
\end{figure*}

In Kepler-88, the expected chopping period is $P_\mathrm{chop} = 21.5$ days.\footnote{The periods $P_b$ and $P_c$ used in determining the expected chopping signal are from the linear ephemeris (as in Section \ref{sec:keplerian}, which are substantially different from the values used in intializing a N-body fit to the TTVs.}  We identified the chopping signal by fitting the \citet{Holczer2016} TTVs with a high-order polynomial\footnote{We used the lowest-degree polynomial that removed significant peaks at much longer periods than the expected synodic chopping signal} (degree 18) and subtracted this polynomial fit from the TTVs (Figure \ref{fig:koi142-ttv}, middle panel, blue points).  The high-frequency variations in the TTVs show a characteristic chopping signal.  Note that the strength of the chopping comes and goes at different phases of the TTV super-period.  This episodic strength of the chopping signal is well-reproduced by our best-fit model (orange crosses).

\begin{figure}
    \centering
    \includegraphics[width=0.5\textwidth]{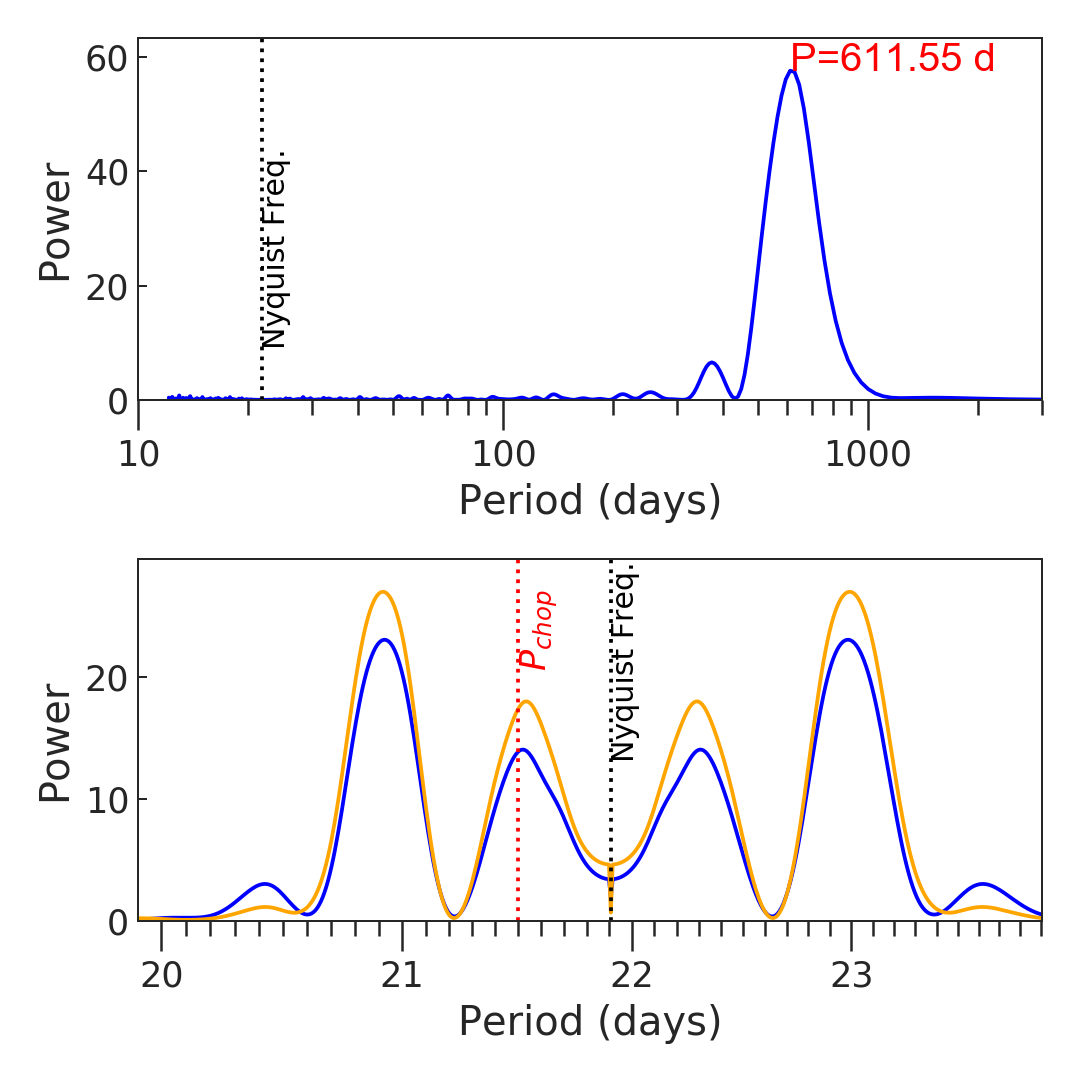}
    \caption{Top: The Lomb-Scargle periodogram of the TTV signal has a peak at 611 days, the super-period of the Kepler-88 TTVs.  Bottom: The Lomb-Scargle periodogram of the chopping signal (after removing the TTV super-period) of both the observed TTVs (blue) and our best-fit model (orange).  The predicted chopping signal is at 21.5 days (below the Nyquist frequency), but the strongest frequency is at 20.9 (or 23) days.  The peak at 20.9 days days corresponds to the 2:3 mean motion resonance of planets b and c.}
    \label{fig:chop_period}
\end{figure}

We used a Lomb-Scargle periodogram to identify the super-period of the TTVs at 611 days (Figure \ref{fig:chop_period}, top).  After we removed the high-degree polynomial, the Lomb-Scargle periodogram of the TTV residuals had its highest peak at P = 20.9 or 23 days (which is the mirror reflection of 20.9 days about the Nyquist frequency), and and a peak at the expected chopping signal at 21.5 days (bottom).  Note that peak periods are reflected about the Nyquist frequency, $1/(2\bar{P_b}) = 1/21.9$ days$^{-1}$.

\begin{figure}
    \centering
    \includegraphics[width=0.5\textwidth]{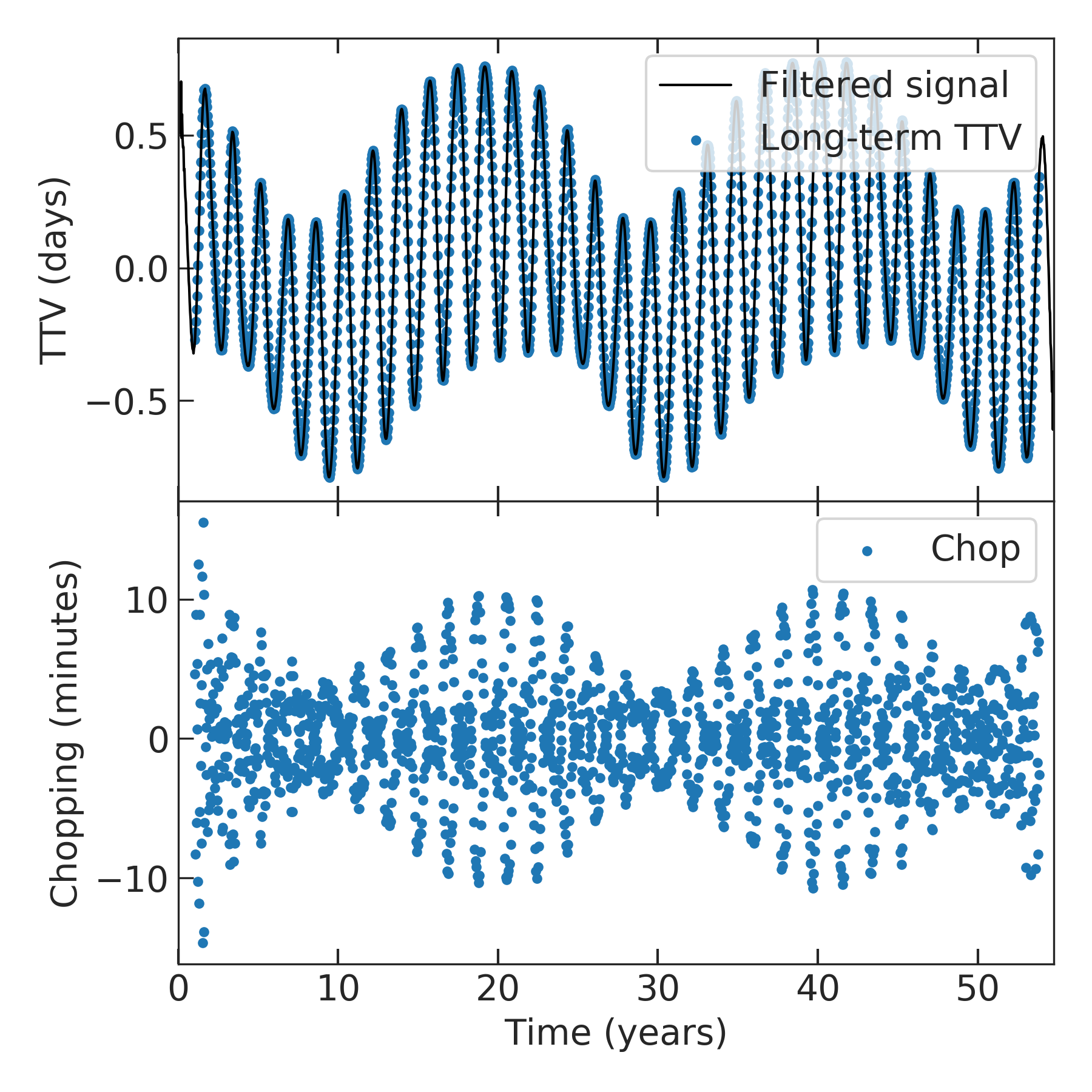}
    \caption{Top: Long-term TTVs of Kepler-88 b predicted from our best-fit model (blue points) and a Fast Fourier Transform (FFT) of to the TTVs (black line).  In addition to the 600-day super-period, there is a 20 year super-super period.  Bottom panel: The long-term chopping signal is computed by subtracting the FFT from the TTVs in the upper panel.  The chopping signal amplitude varies on the timescales of the TTV super-period and super-super period.}
    \label{fig:longchop}
\end{figure}

\begin{figure}
    \centering
    \includegraphics[width=0.5\textwidth]{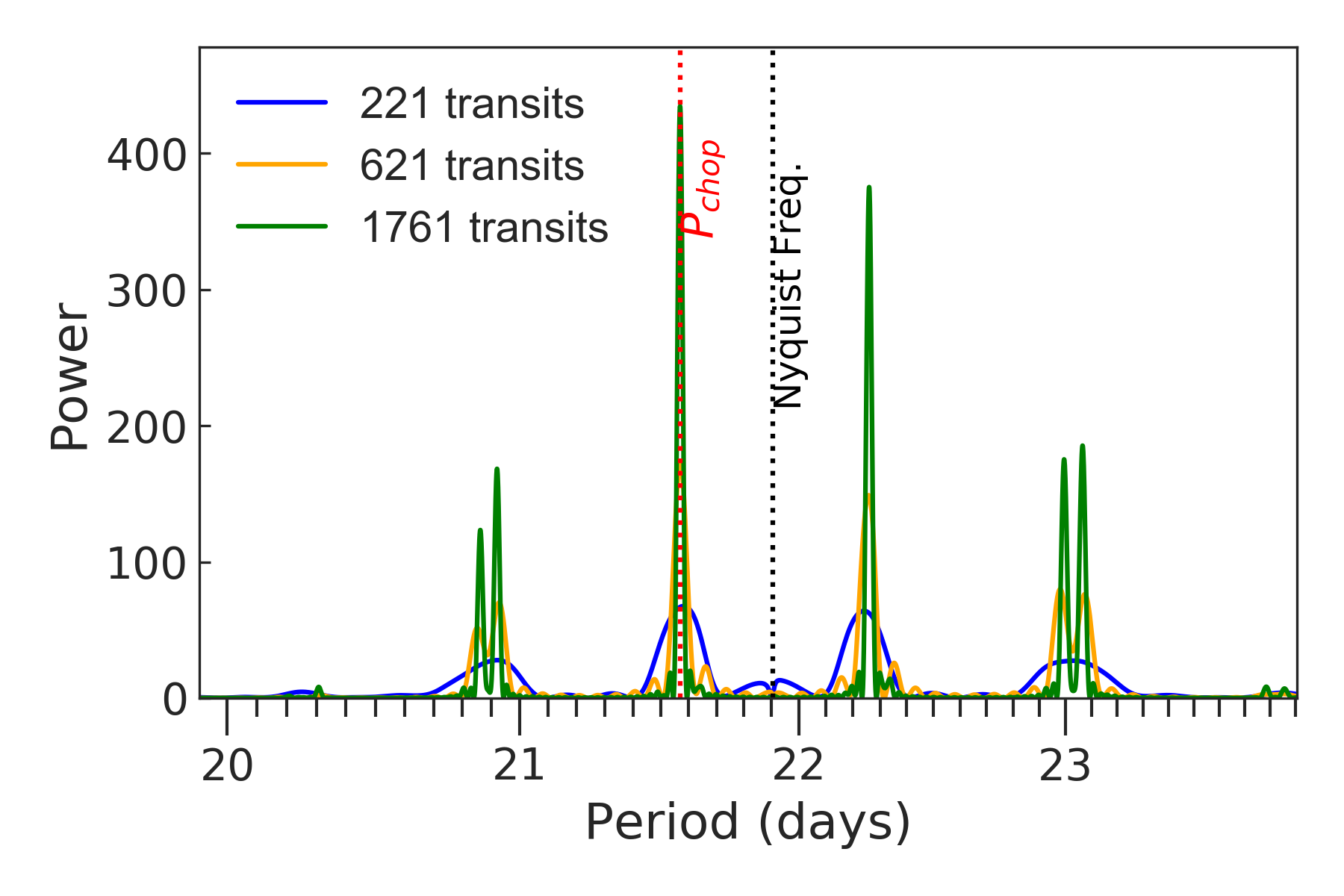}
    \caption{Lomb-Scargle periodogram of the long-term chopping signal from Figure \ref{fig:longchop} based on the middle 221 (blue), 621 (orange), or 1761 (green) consecutive transits.  With only 221 transits, the peaks at 21.5 days (the synodic chopping signal), its reflection about the Nyquist frequency, and the peaks 20.9/23 days are visible.  With additional transits, the power at each of these peaks grows.  The peaks at 20.9/23 days split, but the peaks at the synodic chopping period and its Nyquist reflection do not.}
    \label{fig:longchop_pgram}
\end{figure}

To better understand the origin of the peak at 20.9 or 23 days, we examined the behavior of the TTV chopping signal on longer timescales.  We simulated TTVs for 50 years using the best-fit parameters (Figure \ref{fig:longchop}, top panel).  The long-term TTVs have both the a super-period at 611 days, and a super-super period at 20.5 years (7500 days).  We used a fast Fourier transform to construct a low-pass filter and subtracted the filter from the model, thus obtaining the chopping signal (bottom panel).  In Figure \ref{fig:longchop_pgram}, we show the Lomb-Scargle periodogram of the chopping signal based on different numbers of consecutive transits.  A periodogram of the 221 transits from the middle of the simulated chopping signal yields peaks at the chopping frequency (21.56 days) and 20.9/23 days.  Including more transits increases the power at these distinct periods, and reveals splitting of the 20.9/23 day period.  The peak at 20.91 days is at $1/(2/P_b - 3/P_c)$, i.e., the 2:3 resonance.\footnote{We looked for similar peaks at the other first-order resonances, but only found peaks near j:j-1 for odd values of j.  These peaks also happen to be near the aliases produced by the window function.  Further investigation of the full sequence and dynamical origin of the chopping signal periodogram is outside the scope of this paper.}  The splitting appears consistent with the super-super period, as the peaks occur at 20.91 and $20.86 = 1/(1/20.91 + 1/7500)$ days.

\begin{figure}
    \centering
    \includegraphics[width=0.5\textwidth]{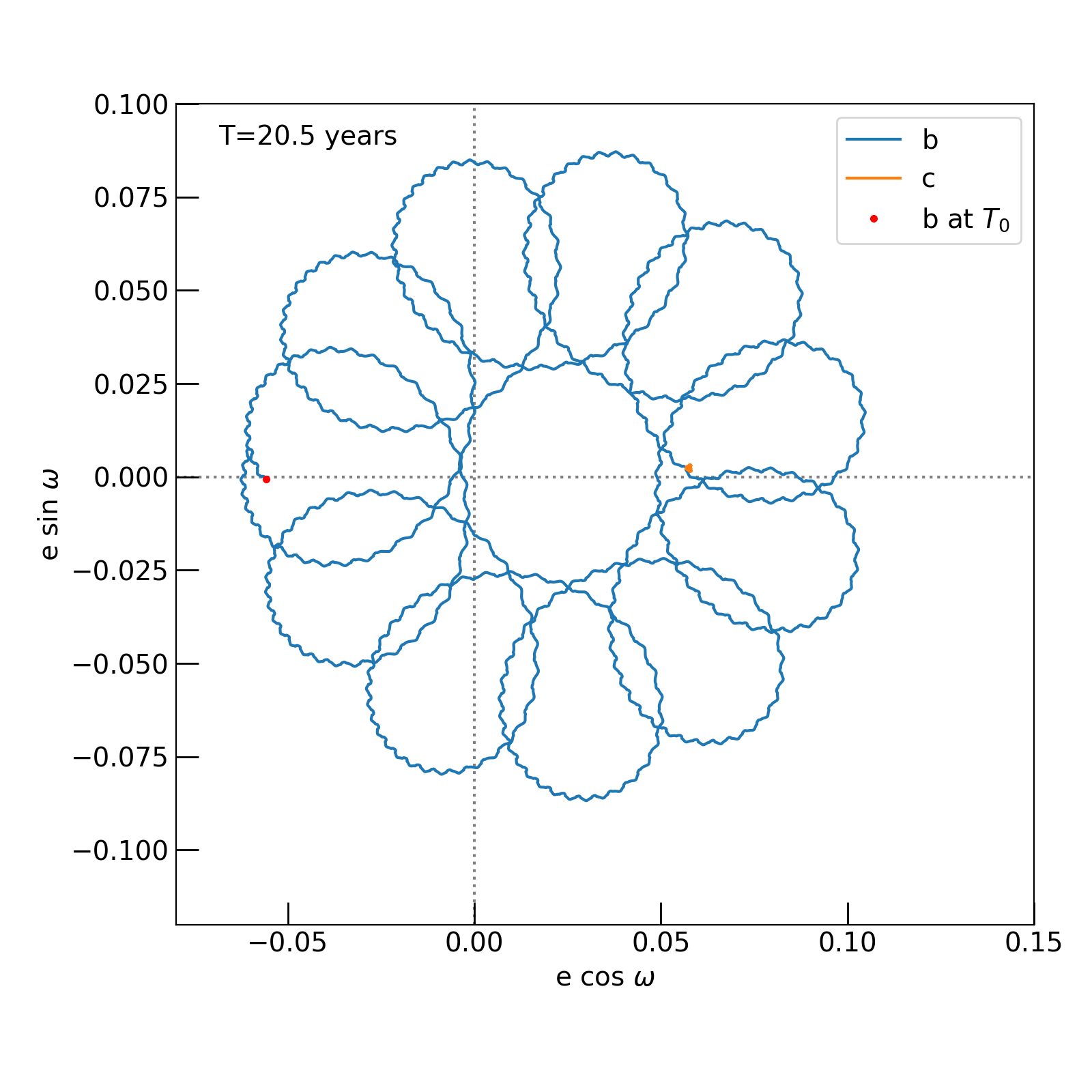}
    \caption{Parametric evolution of \ecosw, \esinw, in a simulation with just the inner two planets.  In 20.5 years, planet b traces a flower/spirograph pattern through \ecosw, \esinw\ space.  Each epicyclic petal is traced in 611 days (the super-period of the TTVs), and the full flower is completed in 20.5 years (the super-super period of the TTVs).  Note that the precession of the super period and the super-super period have opposite signs.  The small zig-zags are at the timescale of synodic chopping.}
    \label{fig:flower}
\end{figure}

To investigate the origin of the super-super period, we plotted the evolution of \ecosw\ and \esinw\ in a simulation with just the inner two planets (Figure \ref{fig:flower}), and confirmed that the super-super period (20.5 years) is the timescale of precession of planet b due to planet c.\footnote{The longer timescale interactions of planet d are discussed in \S\ref{sec:longterm}.}  There are approximately 11 retrograde epicycles during the 20.5 year precession; each of these epicycles corresponds to a TTV super-period of $\sim611$ days.

\section{Photodynamical Fit to Transits and RVs}
A photodynamical fit is produced by optimizing an N-body model to fit photometry (and in this case also RVs).  Unlike a simultaneous fit to TTVs + RVs, the photodynamical fit must reproduce the transit time and also the transit depth, duration, and shape at each epoch.  It is computationally more expensive than a fit to TTVs, but also potentially more informative, as it enables an exploration of the inclinations ($i$) and longitudes of ascending node ($\Omega$) of the planets, which can be constrained by the transit depths and durations.

To improve upon the RV + TTV solution, we used an iterative photodynamical forward-model to simultaneously fit the photometry and RVs of the Kepler-88 system.  We used the code \texttt{Phodymm} which has previously been used to model and fit photometry from the \Kepler\ prime mission in Kepler-223, \citep{Mills2016}, Kepler-444 \citep{Mills2017_K444} and Kepler-108 \citep{Mills2017_K108}, and the combined \Kepler\ prime photometry and Keck-HIRES RVs in Kepler-25, Kepler-65, and Kepler-68 \citep{Mills2019_giants}.  \texttt{Phodymm} is a Runge-Kutta N-body integrator that can simultaneously forward-model photometry and RVs for N planets and one star.  The transit shape is reproduced with the prescription given in \citet{Pal2011}.  This model includes a transit shape described by \citet{Mandel2002}, with the quadratic limb-darkening coefficients of \citet{Claret2000}.  For simplicity, \texttt{Phodymm} assumes that the velocity of the planet is constant during transit.  For input parameters, it can accept Cartesian, asterocentric, or Jacobi coordinates.  We used the Jacobi orbital elements: orbital period $P$, time of conjunction $T_c$, eccentricity $e$, inclination $i$, longitude of ascending node $\Omega$, and argument of periastron passage $\omega$ all of which were defined at epoch BJD $= 2454954.62702$.  Additional input parameters were the planet-to-star radius ratio $\rpl/\rstar$, and planet mass $\mpl$ for each planet, as well as the stellar mass \mstar, radius \rstar, dilution $D$, and quadratic limb-darkening coefficients $c_1$ and $c_2$.

\subsection{Photometry}
We downloaded the photometry of Kepler-88 (KOI-142, KIC 5446285) obtained during the Kepler prime mission from the MAST archive \footnote{https://archive.stsci.edu/}.  Where available (quarters 4-17), we used short cadence data; we used long cadence data elsewhere.  We detrended the photometry in the manner of \citet{Mills2017_K108}.  First, we segmented the lightcurve into chunks of approximately one day, masking any transits within each chunk.  We then fit the photometry in each chunk with a cubic polynomial to model the continuum, including both systematic effects and stellar rotation.  We divided the observed flux by our continuum model to obtain normalized photometry.  We multiplied all the uncertainties by a scale factor such that the out of transit reduced $\chi^2$ is 1.0, for both long cadence and short cadence independently.

\subsection{Photodynamical Fit \label{sec:photodynamical}}
Since no stellar companions are known, we fixed the transit dilution at zero for all the planets, and we fixed planet-to-star radius ratios of the two non-transiting planets, the longitude of ascending node for planet b of $\Omega_b=0.0$ (since this is an arbitrary angle on the sky plane), and the inclination and longitude of ascending node for planet d at $i_d = 89^\circ$, $\Omega_d = 0.0$ (the RVs only give \msini\ information for this planet, and the TTVs do not help constrain its inclination).  All other parameters were varied.

We arrived at our best estimate of the dynamical parameters in the following manner.  First, we used the parameters published in N13 in conjunction with the software package \texttt{TTVFast} \citep{Deck2014} to minimize our fit to the long-cadence determined TTVs reported in \citet{Holczer2016}.  When our fit to the long-cadence TTVs was optimized, we used our best fit as input orbital elements for \texttt{Phodymm}.  We then ran 40 differential evolution MCMC (DE-MCMC) chains $10^6$ steps each to obtain improved values and formal uncertainties for each of the \texttt{Phodymm} variables.  The chains were well-mixed, based on both a visual inspection of the chain for each parameter and a maximum Gelman-Rubin statistic of 1.05 \citep{Gelman1992}.  Our best-fit parameters and uncertainties for the photodynamical N-body model are in Table \ref{tab:phodymm}.

\begin{figure}
\centering
\includegraphics[width=0.5\textwidth]{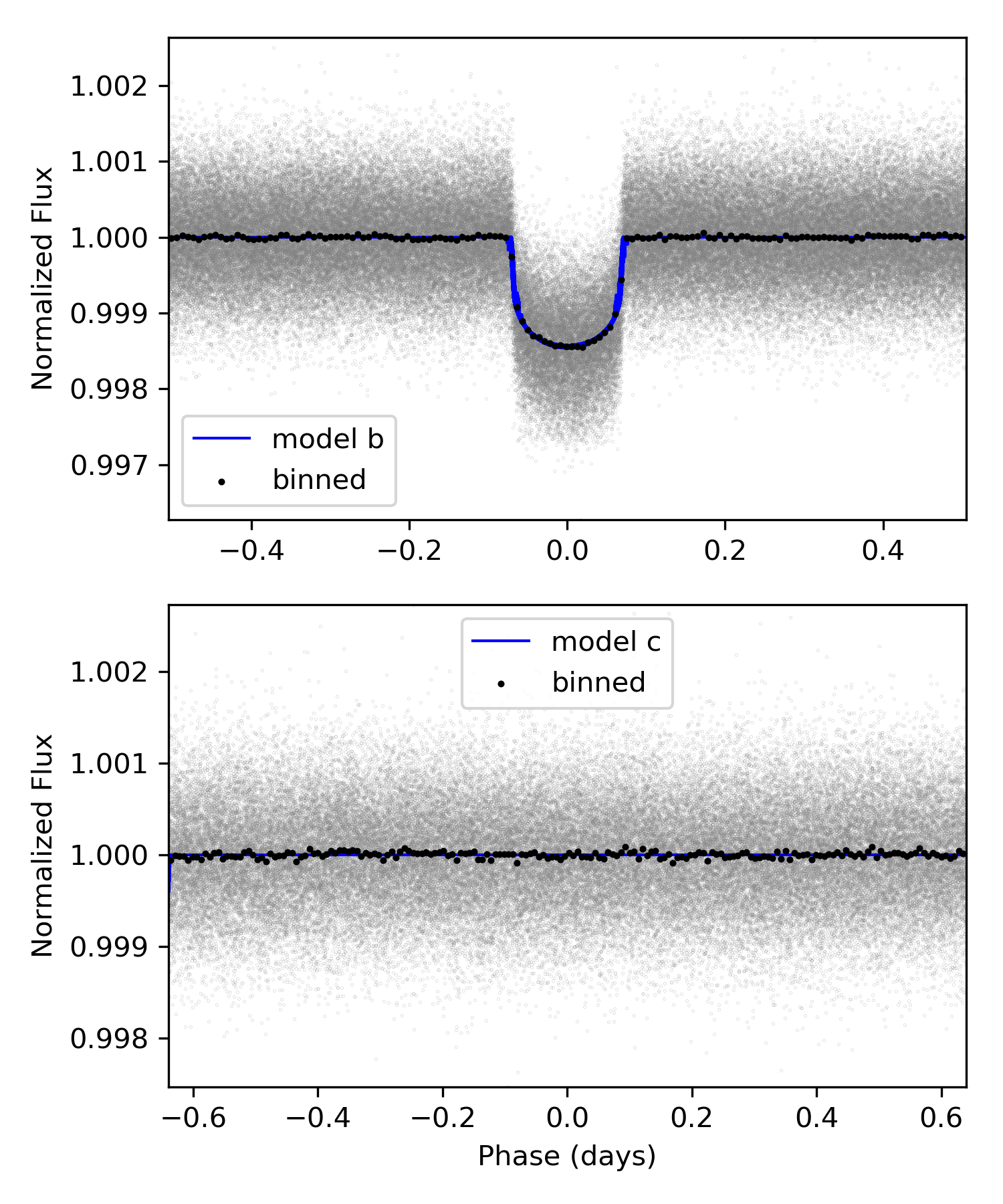}
\caption{The long and short cadence \Kepler\ photometry (gray) has been phase-folded to the best-fit times of conjunction of planets b (top) and c (bottom).  The black points are a running median to more clearly show the transit shape, and the blue curve is the phase-folded model.  For planet b, the transit ingress and egress are sharp, with no evidence of horizontal smearing from improperly determined transit times.  Transits of planet c are not detected.}
\label{fig:koi142-phasefold}
\end{figure}

\begin{figure}
\centering
\includegraphics[width=0.5\textwidth]{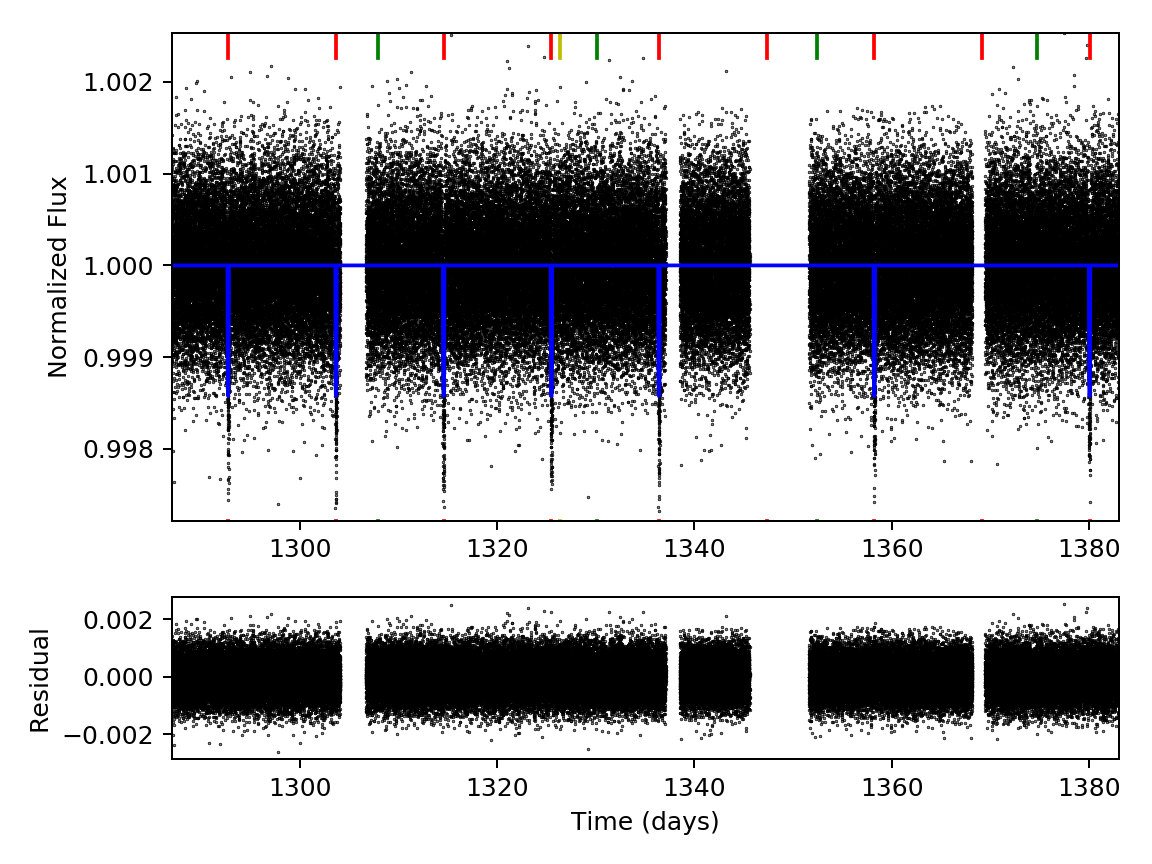}
\caption{The short cadence \Kepler\ photometry (black) within the 2-$\sigma$ confidence interval of the expected conjunction of planet d (the maximum likelihood value for which is marked with the yellow notch at the top of the figure).  Several transits of planet b (red notches) and non-transiting conjunctions of planet c (green notches) also occur during this time.  The best-fit photodynamical model is the blue line.  We do not visually detect a transit of planet d in these data.}
\label{fig:photometry-zoom}
\end{figure}

Figure \ref{fig:koi142-phasefold} shows the photometry phase-folded to the individual transit times of Kepler-88 b.  The sharp ingress and egress indicate that the individual  transit times have been well-determined.  Furthermore, the distribution of the photometric residuals during transit are identical to the distribution of the photometric residuals outside of transit, and both are Gaussian, with a standard deviation of 550 parts per million per exposure.  The residuals do not have strong correlated features, suggesting that the individual transit times, depths, shapes, and durations have been well-modeled.\footnote{We tested this assertion by computing the auto-correlation function of the residuals, the magnitude of which did not exceed 0.002 for lags larger than unity.}  No transits of planets c or d were detected: even a grazing transit of a 1\,\rjup\ planet with impact parameter $b=\rstar$ would produce a deeper transit than that of planet b, which would be easily identifiable in the photometry of Figure \ref{fig:photometry-zoom}.

\begin{deluxetable}{lll}




\tablecaption{Phodymm MCMC Posteriors \label{tab:phodymm}}


\tablehead{\colhead{Parameter} & \colhead{Units} & \colhead{Median $\pm1\sigma$}}
\startdata
Period$_b^\dagger$                     & days                 & $10.91647\pm0.00014$\\
T$_0,_b$                       & BJD$-$BJD$_0$        & $55.08069\pm0.00061$\\
$\sqrt{e}\cos\omega_b$         &                      & $-0.23578\pm0.00031$\\
$\sqrt{e}\sin\omega_b$         &                      & $0.0044\pm0.0027$   \\
i$_b$                          & $^\circ$             & $90.97\pm0.12$      \\
M$_{jup,b}$                   & Jup                  & $0.0300\pm0.0036$     \\
R$_b$/R$_s$                  &                 & $0.03515\pm0.00018$ \\
Period$_c$                     & days                 & $22.26492\pm0.00067$\\
T$_0,_c$                       & BJD$-$BJD$_0$        & $61.353\pm0.025$    \\
$\sqrt{e}\cos\omega_c$         &                      & $0.2392\pm0.00095$  \\
$\sqrt{e}\sin\omega_c$         &                      & $-0.0044\pm0.0033$  \\
i$_c$                          & $^\circ$             & $93.15\pm0.68$      \\
$\Omega_c$                     & $^\circ$             & $-0.43\pm0.19$      \\
M$_{jup,c}$                   & Jup                  & $0.674\pm0.016$     \\
Period$_d$                     & days                 & $1403\pm14$     \\
T$_0,_d$                       & BJD$-$BJD$_0$        & $1335\pm19$     \\
$\sqrt{e}\cos\omega_d$         &                      & $0.63\pm0.03$     \\
$\sqrt{e}\sin\omega_d$         &                      & $0.08\pm0.05$     \\
Msin$i_{jup,d}$                   & Jup                  & $3.05\pm0.16$       \\
M$_s$                          & solar                & $0.990\pm0.023$      \\
R$_s$                          & solar                & $0.897\pm0.016$     \\
c$_1$                          &                      & $0.394\pm0.062$     \\
c$_2$                          &                      & $0.292\pm0.096$     \\
\hline
M$_b$                          & Earth                & $9.5\pm1.1$         \\
M$_c$                          & Earth                & $214.1\pm5.2$       \\
Msin$i_d$                          & Earth                & $965\pm44$      \\
R$_b$                          & Earth                & $3.438\pm0.075$     \\
$\rho_b$                       &                      & $1.29\pm0.16$       \\
$e_b$                          &                      & $0.05561\pm0.00013$ \\
$e_c$                          &                      & $0.05724\pm0.00045$ \\
$e_d$                          &                      & $0.41\pm0.03$     \\
$I_{bc}$                       & $^\circ$             & $2.23\pm0.62$\\
\enddata


\tablecomments{The MCMC parameters are above the line; derived parameters are below the line. All parameters are computed at epoch $T_{0,\mathrm{BJD}} = 2454954.62702$.  $\Omega_b$ is an arbitrary reference angle and was fixed at 0.0.  The Photodynamical solution was not sensitive to the inclination of planet d, which we fixed at $i_d = 89^{\circ}$.}
\tablenotetext{\dagger}{Not the same as the linear ephemeris, which is $\bar{P_b} = 10.95$ days.}


\end{deluxetable}

\subsection{Transit Times}
One outcome of the photodynamical modeling is a model-dependent determination of the individual transit times.  The transit midpoint times T$_i$, impact parameters b$_i$, and planet velocities during transit v$_i$, and their uncertainties, are given in Table \ref{tab:tbv00_01} for Kepler-88 b from the date of the first \Kepler\ transit through November 2022.

\begin{deluxetable*}{ccccccc}




\tablecaption{Kepler-88 b Transit Times and Velocities \label{tab:tbv00_01}}

\tablehead{\colhead{Epoch} & \colhead{T$_i$} & \colhead{T$_i$ Err.} &\colhead{b$_i$}  &\colhead{b$_i$ Err.}& \colhead{v$_i$} & \colhead{v$_i$ Err.}\\ 
 \colhead{} & \colhead{(BJD$-$BJD$_0$)} & \colhead{(BJD$-$BJD$_0$)} & \colhead{(AU)} & \colhead{(AU)}& \colhead{(AU/day)} & \colhead{(AU/day)} } 

\startdata
0 & 55.0801 & 0.0006 & 0.0017 & 0.0002 & 0.0554 & 0.0004 \\
1 & 66.0069 & 0.0006 & 0.0017 & 0.0002 & 0.0554 & 0.0004 \\
2 & 76.9294 & 0.0006 & 0.0017 & 0.0002 & 0.0556 & 0.0005 \\
3 & 87.8535 & 0.0006 & 0.0016 & 0.0002 & 0.0556 & 0.0005 \\
4 & 98.7724 & 0.0006 & 0.0016 & 0.0002 & 0.0558 & 0.0005 \\
5 & 109.6940 & 0.0006 & 0.0016 & 0.0002 & 0.0559 & 0.0005 \\
\enddata


\tablecomments{T$_i$ refers to the transit midpoint time, b$_i$ is the impact parameter, and v$_i$ is the planet velocity during the $i^{\mathrm{th}}$ transit.  BJD$_0=$2454900.  The full table is available in machine readable form.  The first few lines are shown here for content and format.}


\end{deluxetable*}

\section{Results}
\subsection{Confirmation of a giant planet near a 2:1 MMR}
In both our RV-only and our photodynamical analysis, we confirm the existence of a giant planet at \Pcavgshort\ with a mass of 200\,\mearth.  A Lomb-Scargle periodogram of our RVs produces a very strong peak at 22.26 days, with no significant peaks at aliases or harmonics of this period, indicating that 22.26 days is in fact the period of the perturbing giant planet (Figure \ref{fig:koi142-pgram}).  
\subsection{Discovery of a long-period giant planet}
\begin{figure}[ht!]
\centering
\includegraphics[width=0.5\textwidth]{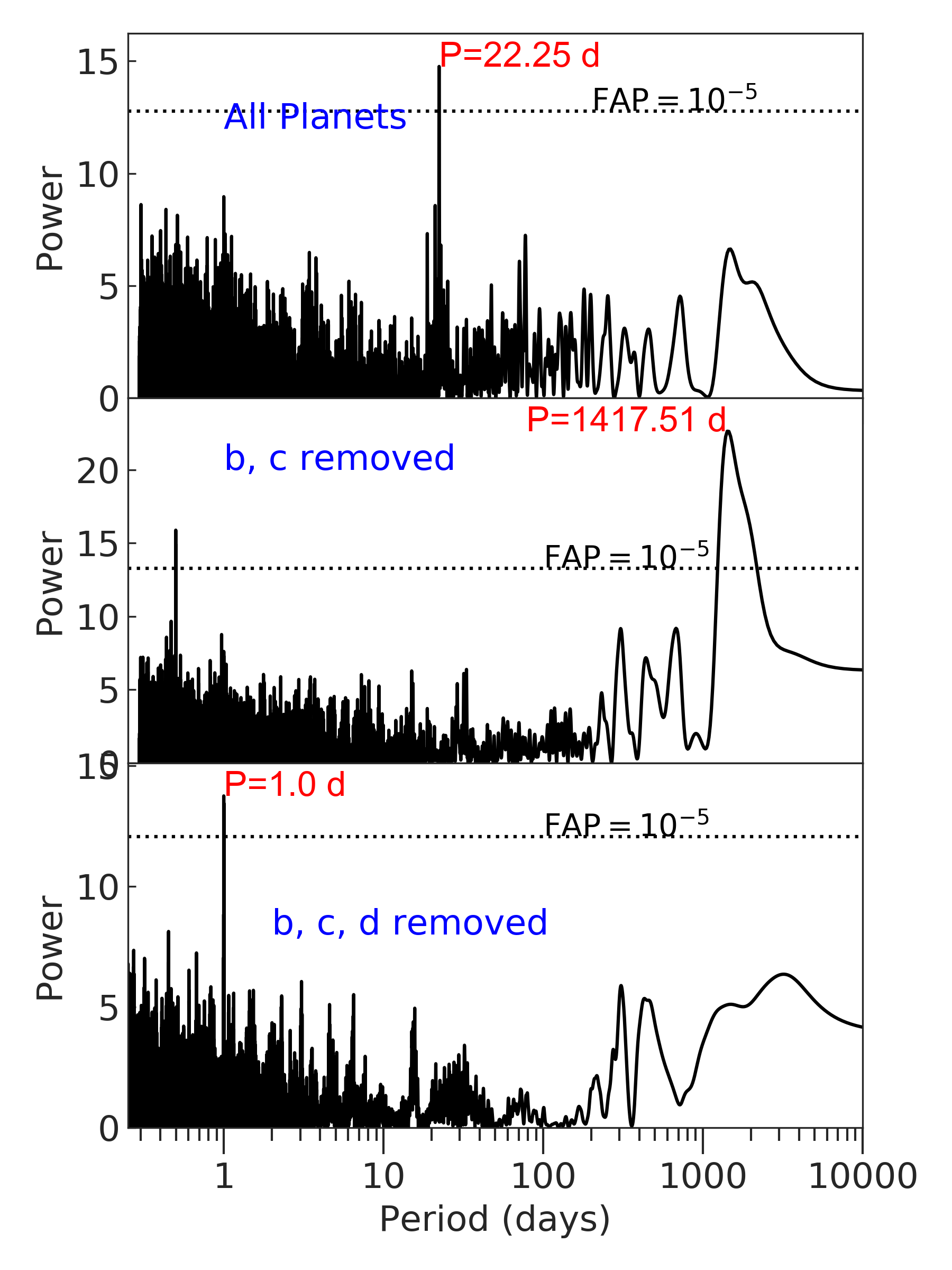}
\caption{Top: Lomb-scargle periodogram of the Kepler-88 RVs.  Middle: Lomb-scargle periodogram to the residual RVs, after subtracting the best-fit N-body two-planet model (planets b \& c).  The significant peak at 1400 days is strong evidence for a third planet in the system.  Bottom: Lomb-scargle periodogram to the residual RVs, after subtracting the best N-body three-planet model.  There is a strong peak at 1.0 days that is likely the consequence of correlated noise and our window function.  If there is a fourth planet, its orbital period is not yet apparent.  The spikes near $P=1$ days and higher-frequency harmonics are aliases of the long-period planet(s) and/or long-term RV noise.  The false alarm probabilities (FAP) are computed by bootstrap resampling the RVs.}
\label{fig:koi142-pgram}
\end{figure}

In the Keck-HIRES RV data, we identify a third planet at \Pdval\ with \mplsini=\mdval.  When we compute the Lomb-Scargle periodogram to the residual RVs of a 2-planet fit (where the orbits are N-body), there is a significant peak at $P=1413$ days, and there are no other peaks with comparable power (see Figure \ref{fig:koi142-pgram}). We find evidence for the third planet in the significantly improved $\chi^2$ statistic to the RVs, which we summarize in Table \ref{tab:bic_compare}.  Without the third planet, our best N-body fit to the TTVs + RVs has $\chi_{\rm{RV}}^2 = 106.2$ (see Figure \ref{fig:koi142-2pl}).  Including the third planet near $P=1413$ days results in $\chi_{\rm{RV}}^2 = 51.6$ (see Figure \ref{fig:koi142-3pl}).  The inclusion of the third planet substantially improves the fit to the RVs while simultaneously reducing the HIRES RV jitter by a factor of $\sim3$.  However, the goodness of fit to the TTVs does not change significantly between the 2-planet and 3-planet models, indicating that the TTVs provide essentially no evidence for the existence of planet d.  This is unsurprising, since at \Pdval, the outer planet causes negligible changes to the orbits of the inner planets on the timescale of the \Kepler\ baseline.  Therefore, the outer planet was only detected in RVs.  The RV-led discovery highlights the importance of multi-method follow-up of the most architecturally interesting \Kepler\ planetary systems.

\subsection{Comparison of the RVs Only vs. Photodynamical Model}
The photodynamical fit provides some advantages over the RVs alone.  Although the RV alone identifies the period of planet c as $P_c=\PcRV$, the photodynamical fit tightens the uncertainty by almost an order of magnitude, finding $P_c=\Pcval$.  Note that the precision on our determination of the orbit of planet c is a factor of 3 better than that of N13, which found $22.3397^{+0.0021}_{-0.0018}$ days; the improvement in the precision of the orbital period must come from the additional \Kepler\ photometry in our analysis, since the RVs alone did not determine the orbit of planet c as precisely as the TTV-based N13 work.  Also, the RVs are only able to provide a mass upper limit for planet b, but the planet's period, radius, and mass are determined with high confidence in the photodynamical analysis: $P_b = \Pbval$ (at epoch BJD$=2454954.62702$), $R_b = \Rbval$, $M_b = \Mbval$.  The superior performance of the photodynamical model for the inner planets illustrates the complementary nature of transit photometry and radial velocities: together, these techniques reveal more about the 3D architecture of a planetary system than each of these techniques does alone.  The superior mass determination of planets b and c in the photodynamical model can be traced to the chopping signal in the TTVs. 

\subsection{Architectural constraints from photodynamical modeling}
From the photodynamical analysis, we determined that the two inner planets, b and c, are apsidally anti-aligned.  Our result agrees with N13.  Apsidal anti-alignment is a predicted outcome of convergent or divergent Type I migration in a viscous disk \citep[][and references therein]{Nelson2018}.  The combination of the near-resonant configuration for the inner planets and their apsidal anti-alignment could suggest a history of migration and resonant trapping, although the current anti-alignment is likely a short-lived coincidence (see Figure \ref{fig:flower}).

Planet d's longitude of periastron passage is nearly aligned with that of planet b, and anti-aligned with that of planet c.  The apsidal alignment of planet d is far less meaningful, since it is dynamically decoupled from the inner two planets.  However, the eccentricity of planet d is large ($e_d = \edval$) compared to the inner planets ($e_{{b,c}} \approx 0.06$).  The high eccentricity of planet d could be explained by planet-planet scattering or Kozai oscillations.  By contrast, the modest eccentricities of planets b and c could possibly be explained by the equilibrium of disk and/or tidal circularization and N-body eccentricity pumping.

Because both the inclinations and relative longitude of ascending node are constrained, we can compute the mutual inclination between planets b and c:
\begin{equation}
\begin{split}
\mathrm{cos}(I_{bc}) & = \mathrm{cos}(i_b)\,\mathrm{cos}(i_c)\\
& + \mathrm{sin}(i_b)\,\mathrm{sin }(i_c)\, \mathrm{cos}(\Omega_b - \Omega_c)\\
\end{split}
\end{equation}
We find that the mutual inclination of planets b and c is tightly constrained: $I_{bc} = 2.23\pm0.62^\circ$.  To test the extent to which the inclinations of the planets are constrained photodynamically, we initialized a DE-MCMC experiment with 40 walkers with our best fit, but flipped the inclination of planet c about the stellar meridian ($i=90^\circ$), allowing the same parameters to vary as in our control trial.  In general, it is difficult for a DE-MCMC exploration to find this parameter space, because inclinations near $90^\circ$ for planet c would produce deep transits, which are not observed.  However, our inclination-flipped experiment performed substantially worse than our best fit, with $\Delta \rm{BIC} = 33$ (see Table \ref{tab:bic_compare}).  Therefore, our findings strongly disfavor the model in which planets b and c are on opposite sides of the star; rather, they seem to be on the same side of the star.

\begin{figure}
    \centering
    \includegraphics[width=0.5\textwidth]{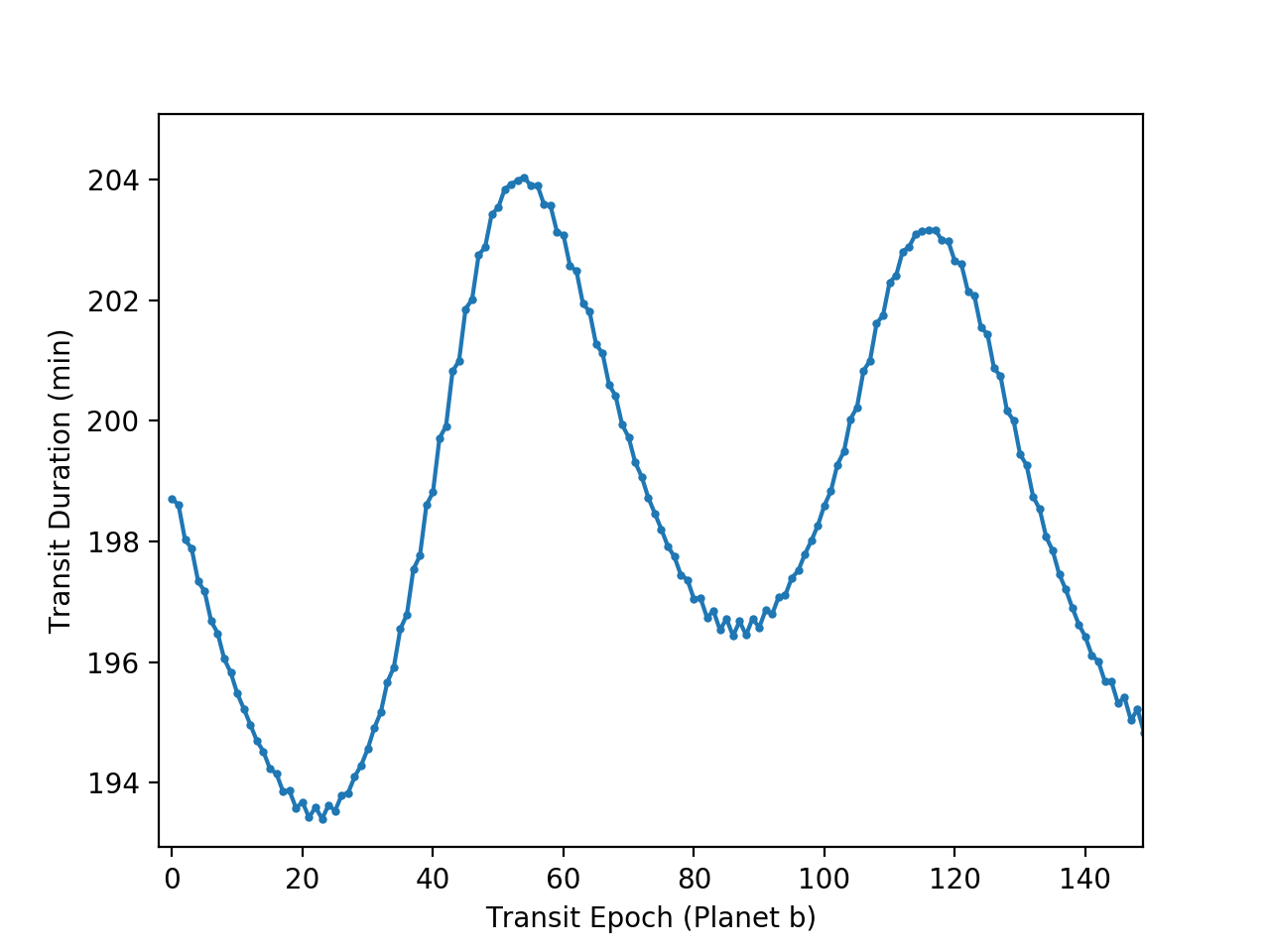}
    \caption{The best-fit durations of the individual transits from our photodynamical model.  There are both long-term TDVs and a chopping signal.}
    \label{fig:TDVs}
\end{figure}

Constraints on the mutual inclinations of planets likely come from transit duration variations (TDVs).  The best-fit solution to our photodynamical model includes substantial TDVs for planet b (see Figure \ref{fig:TDVs}), though these are dominated by in-plane eccentricity precession \citep[i.e.,][]{Nesvorny2013}.  We determined the TDVs based on the velocities and impact parameters in Table \ref{tab:tbv00_01}:
\begin{equation}
T_{i,\mathrm{dur}} \approx \frac{2 \sqrt{(\frac{\rstar}{\mathrm{AU}})^2 - (\frac{b_i}{\mathrm{AU}})^2}}{v_{i,\mathrm{Pl}}\times{\mathrm{day/AU}}} \mathrm{days}
\end{equation}
where $\rstar/\mathrm{AU}$ is the stellar radius in units of AU, $b_i$ is the $i^{\mathrm{th}}$ transit impact parameter in units of AU, and $v_{i,\mathrm{Pl}}$ is the velocity of the planet during the $i^{\mathrm{th}}$ transit in units of AU/day.\footnote{The velocity of the star is ignored here; it is $\mathcal{O}(10^{-5})$ the planet velocity.}  In addition to long-term variation, the TDVs have a chopping signal.

We explored the extent to which we could constrain the inclination of planet d from the TTVs.  By keeping $\msini_d$ constant but varying $i_d$ and $M_d$, we found that the best-fit solution to the TTVs did not significantly degrade.  We tried this experiment in forward-modeling the TTVs with \texttt{TTVFast} and also with \texttt{Phodymm}.  In both cases, a wide range of mutual inclinations between planet d and the inner planetary system are supported by the data.  For example, initializing the inclination of planet d at $i_d=30^\circ$ from the sky plane, i.e. about $60^\circ$ from the inner planets, and thus requiring its mass to be $M_d=6\,\mjup$, only increased (worsened) the BIC of our photodynamical model by 5, which suggests only modestly better performance of a coplanar model.  We consider the long-term orbital stability of such solutions in the next sub-section.  Simulations in similar planetary systems have found that a long-period giant planet is likely to be coplanar with the inner planets, as this configuration is usually stable for longer periods of time than highly mutually inclined geometries \citep{Becker2017}.  In an analysis of the Kepler-88 system, \citet{Denham2019} found that a planet at semi-major axis 2.4 AU with $e = \edval$  would be stable so long as its mass was $< 20\,\mjup$.  At $\msini_d = \mdvaljup$, Kepler-88 d could take on a wide range of inclinations without exceeding 20\,\mjup\ and thereby disrupting the inner system.

\subsection{Long-term Evolution \label{sec:longterm}}
The architecture of a long-period giant planet accompanying two closer-in planets reminded us of the Kepler-56 system. Kepler-56 is a red giant star hosting two coplanar, transiting planets whose orbits are misaligned with respect to the stellar rotation axis, which is determined from the asteroseismic modes of the star \citep{Huber2013}.  Radial velocity monitoring of the system revealed a long-period, non-transiting giant planet with moderate eccentricity \citep{Otor2016}.  Follow-up theoretical work \citep{Gratia2017} suggested that the outer planet can become eccentric due to planet-planet scattering.  In such cases, additional outer planet(s) are likely ejected, leaving the surviving outer planet on an eccentric and inclined orbit. The perturbations ripple to the inner system, not necessarily disrupting it, but possibly causing precession of the orbital plane that periodically misaligns those two planets from the host star's equatorial plane.\footnote{For the Kepler-88 system, another possible consequence could be to leave the resonant libration in an excited state, accounting for the large TTVs.}

\begin{figure}[htbp]
    \centering
    \includegraphics[width=0.5\textwidth]{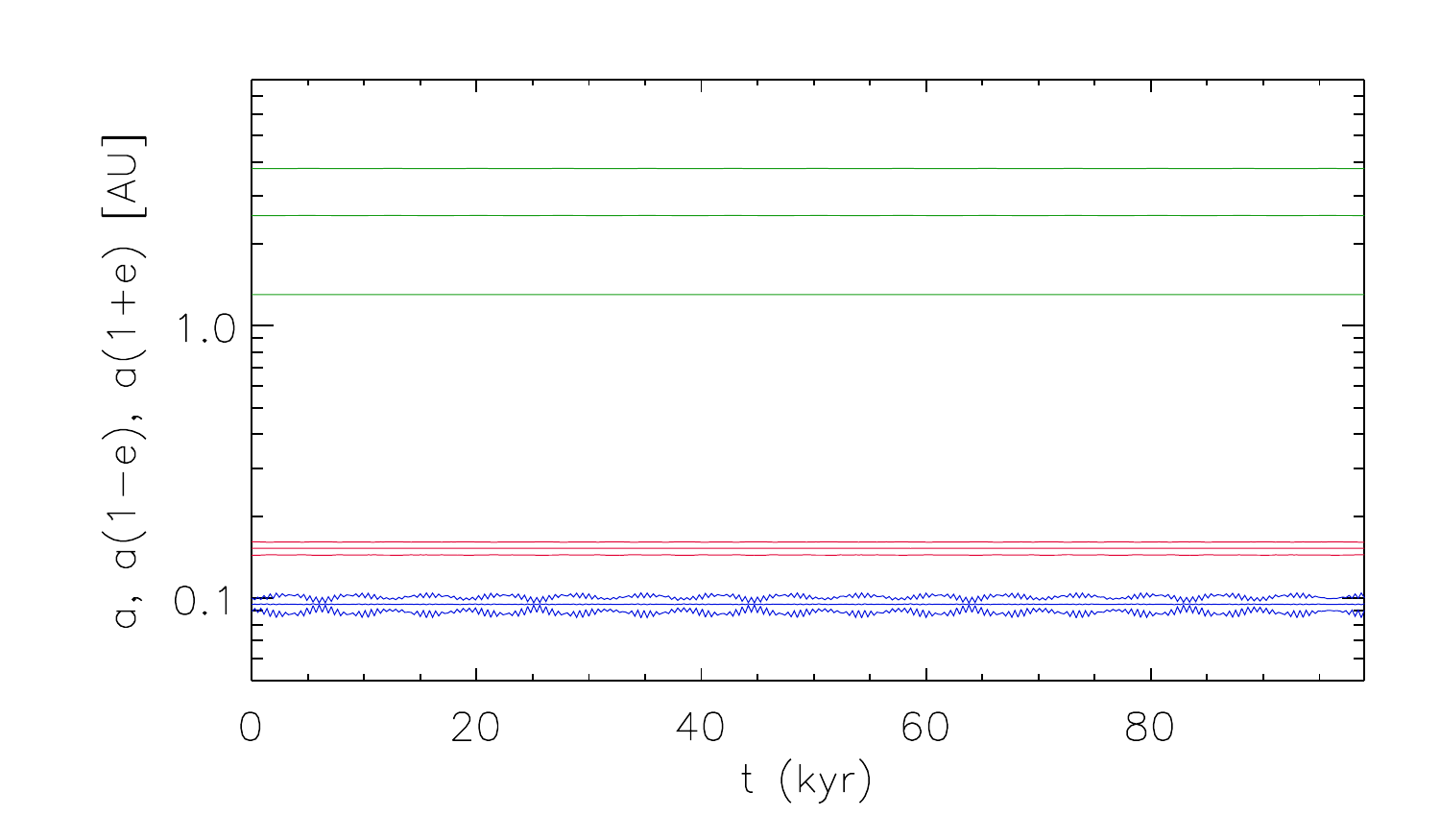}
    \includegraphics[width=0.5\textwidth]{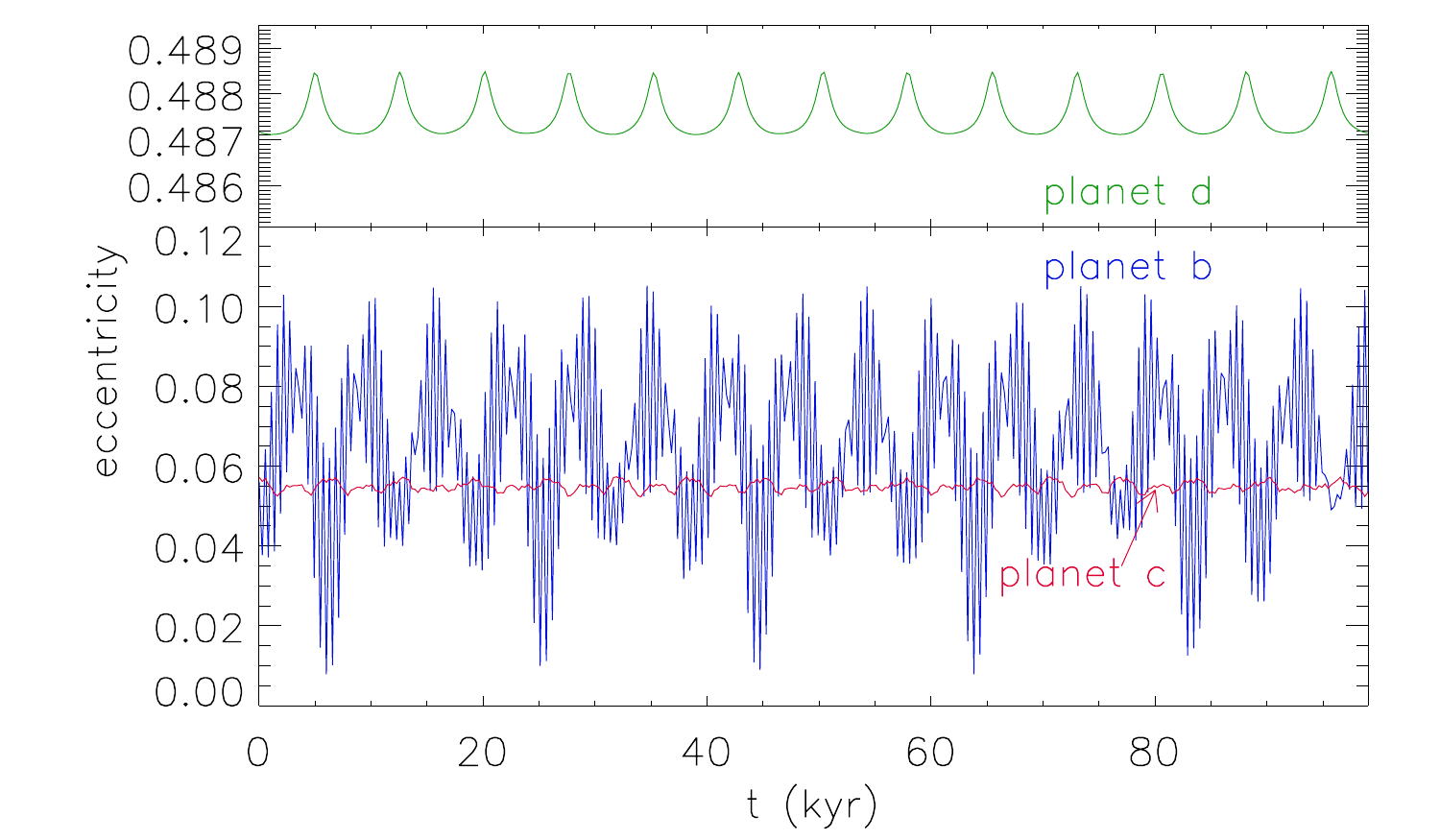}
    \includegraphics[width=0.5\textwidth]{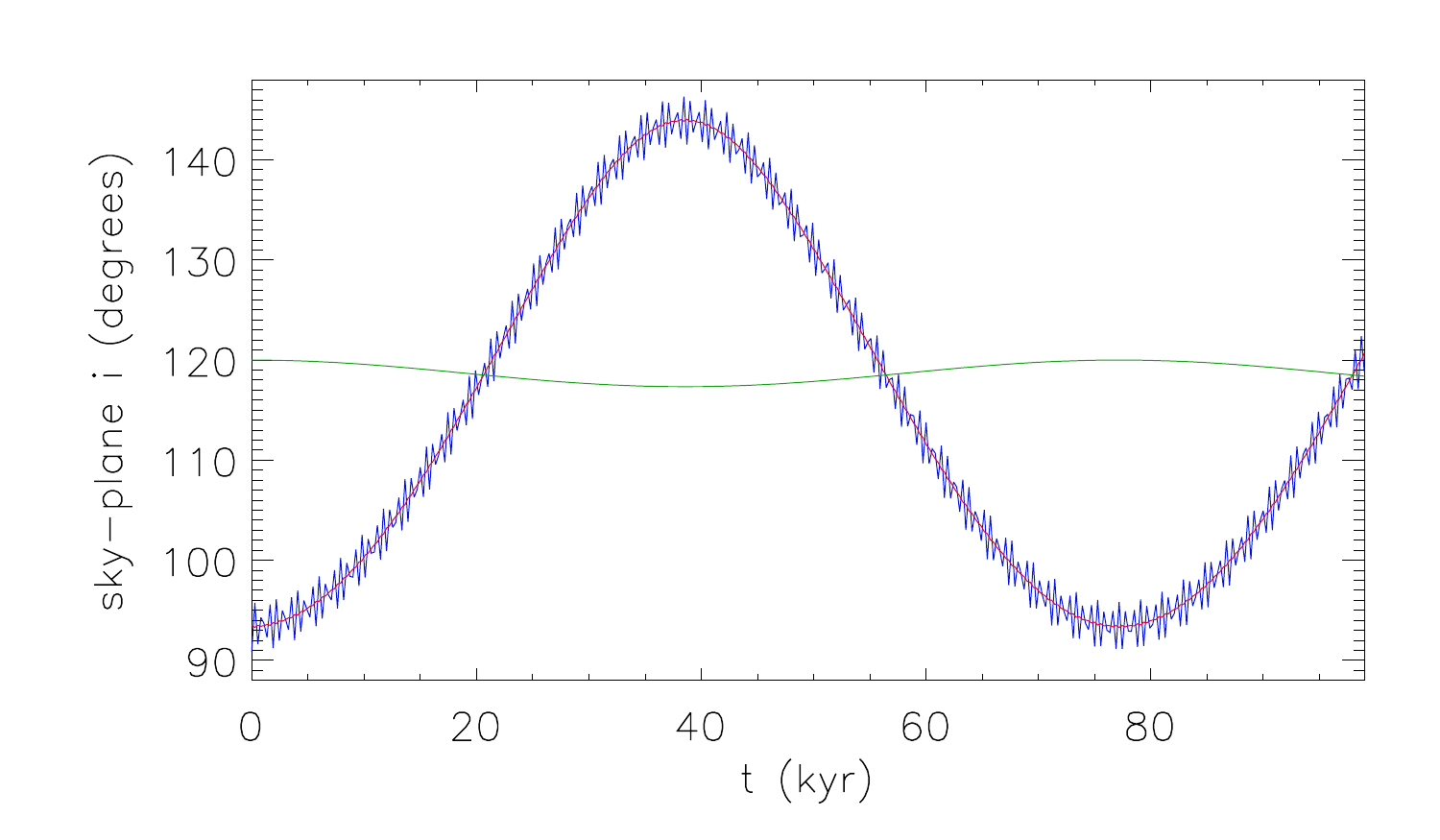}
    \caption{Long-term dynamical simulation of the system.  Planets b, c, and d are represented in blue, red, and green respectively. Top panel: the semi-major axis, periapse distance, and apoapse distance for each planet, as a function of time. Middle panel: Planetary eccentricities. Bottom panel: Inclination to the sky plane.  Over thousands of years, the orbital plane of the inner planets may be torqued through a large angle, away from the transiting configuration.  Note that $i_d$ is unknown.}
    \label{fig:longdynamics}
\end{figure}

Here we run a long-term N-body simulation for Kepler-88 to observe the evolution.  For the simulation we entered a fit of the data into the Mercury package \citep{Chambers1999} and used the Burlisch-Stoer integrator for $0.1$~Myr to record the secular-timescale effects. We have assumed the outer planet is inclined $30^\circ$ from our line of sight and $\sim 27^\circ$ from the inner planets, but that is not constrained from the data. In Figure~\ref{fig:longdynamics} we show that no substantial eccentricity is transferred from the outer giant to the inner planets on these timescales, but a long-term precession effect can excite the inner planets to a large inclination from its original plane, and hence relative to the star. The orbital planes of the two inner planets remain closely aligned to each other. 

We follow \citet{2014BoueFabrycky} to evaluate the secular timescales.  The frequency at which the inner planets would precess due to the outer planet, if the outer planet were not to back-react, is $\nu_3=2.7\times10^{-12}$ rad s$^{-1}$ ($74$~kyr period). The frequency that the outer planet would precess due to the inner planets, if they were not to back-react, is $\nu_4=1.5\times10^{-13}$ rad s$^{-1}$ ($1.33$ Myr period).  Together, the frequency of precession should be $\nu=\cos(I_{cd})(\nu_3+\nu_4)=2.5\times10^{-13}$ rad s$^{-1}$ ($81$-kyr period), where $I_{cd}$ is the mutual inclination between planets $c$ and $d$. The frequencies related to stellar spin precession are several orders of magnitude smaller.  Therefore the inner planets are effectively precessing due to the outer planet, without much back-reaction (as is evident also Figure \ref{fig:longdynamics}, with a precessional frequency of $77$~kyr, near the analytic value), a conclusion that requires only that the outer planet's angular momentum dominates the system, which is true for any possible inclination of the outer planet.  Also, the star's precession cannot keep up with that relatively quick motion. So we expect that a spin orientation measurement of the star, with respect to the transiting planet's orbit, would help diagnose whether the outer planet has a significant inclination with respect to the inner planets.  For instance, if the spin-orbit misalignment is 20 degrees currently, it is likely cycling between 0 degrees and at least 20 degrees, meaning the inclination of planet d with respect to the inner planets is at least 10 degrees. There may, of course, be more to the dynamics than this simple picture, such as a non-trivial scattering history \citep{Gratia2017}.

For completeness, we ran two other simulations for $0.1$~Myr, one in which planet d is nearly coplanar to the inner system, and one corresponding to the a tilt of $60^\circ$ and a doubling of the planetary mass as described above.  Both simulations remained stable, with quasi-periodic oscillations similar to those in figure~\ref{fig:longdynamics}. We conclude that planet d does not threaten the stability of the system for a wide range of inclinations.

\section{Discussion}

\subsection{Implications for planet formation}
Since both planets c and d are gas giants, they must have formed early in the disk lifetime, when gas was abundant.  The presence of multiple giant planets in this system is unsurprising since \feh = \fehval, and the occurrence of giant planets increases with stellar metallicity \citep{Fischer&Valenti2005}. Perhaps additional giant planets were present earlier, or are still present.  Planets c and d likely underwent viscous (Type I) migration in the proto-planetary disk.  As the gas disk dissipated, planet-planet scattering would likely have increased, and low- and high-eccentricity migration likely became important at this time.  The high eccentricity of planet d probably arose due to a significant exchange of angular momentum with another gas-giant planet.

The formation of planet b could have been contemporaneous with the giant planets if the planet were somehow gas-starved, resulting in only a low-mass volatile envelope.  Or perhaps planet b formed when gas was less abundant and was caught in mean motion resonance with planet c during an epoch of inward migration of planet c.

\subsection{Comparison to Other Planetary Systems}
Giant planets are present around a large number of the \Kepler\ systems that host small, transiting planets \citep{Marcy2014, Mills2019_giants}, and perhaps at greater frequency than giant planets occur around field stars \citep{Zhu2018, Bryan2019}.  Kepler-88 joins their ranks.  Furthermore, Kepler-88 has two giant planets.  Other systems with multiple giant planets in addition to small transiting planets include WASP-47 \citep{VanMalle2016} and 55 Cnc \citep{McArthur2004}. The stellar and planetary properties of these systems are summarized in Table \ref{tab:star-compare}.

\begin{deluxetable*}{lcccc}



\tablecaption{Systems with Multiple Giant Planets and Small Transiting Planets \label{tab:star-compare}}

\tablehead{\colhead{Parameter} & \colhead{Units} & \colhead{Kepler-88} & \colhead{55 Cnc} & \colhead{WASP-47}}
\startdata
\sidehead{\bf{Stellar Parameters}}
\teff  & K     & $5466\pm60$     & $5196\pm24$     & $5552\pm75$ \\
\mstar & \msun & $0.99\pm0.024$ & $0.905\pm0.015$ & $1.040\pm0.031$ \\
\rstar & \rsun & $0.897\pm0.016$ & $0.943\pm0.010$ & $1.137\pm0.013$ \\
\feh   & dex   & $0.27\pm0.06$   & $0.31\pm 0.04$  & $0.38\pm0.05$\\
Age    & Gyr   & $1.9\pm1.6$     & $10.2\pm2.5$    & $6.5 \pm 2$ \\	
\sidehead{\bf{Innermost Transiting Planet}}
Letter  &         & b                       & e                     & e \\
Period  & days    &  $10.91649\pm0.00014$   & $0.736539\pm0.000007$ & $0.789592\pm0.000012$\\
\rpl    & \rearth & $3.44\pm0.08$          &  $1.91\pm0.08$        &  $1.810\pm0.027$ \\
\mpl    & \mearth &  $9.5\pm1.1$            & $8.08\pm0.31$	        & $6.83\pm0.66$	\\
\rhopl  & \gcc    & $1.29\pm0.16$           & $6.4\pm0.8$           & $6.35\pm0.64$ \\
Ecc.    &         & $0.05561\pm0.00013$     & $0.040\pm0.027$       & $0.03\pm0.02$ \\
\sidehead{\bf{Innermost Giant Planet}}
Letter  &         & c                       & b                    & b \\
Period & days     &  $22.2649\pm0.0007$    & $14.65152\pm0.00015$ & $4.1591289\pm0.0000042$\\
\msini & \mearth  &  $214.0\pm5.23$          & $264.0 \pm 1.0$      & $363.1\pm7.3	$\\
Ecc.   &          &  \ec                    & $0.0034\pm0.0032$    & $<0.002$\\
\sidehead{\bf{Outermost Known Giant Planet}}
Letter &         & d            &  d              & c \\
Period & days    & $1403\pm14$ & $4825\pm39$     & $588.5\pm2.4$ \\
\msini & \mearth & $965\pm44$   & $1232\pm22$     & $398.2\pm9.3$ \\
Ecc.   &         & \edval       & $0.019\pm0.013$ & $0.296\pm0.017$\\
\enddata
\tablecomments{For Kepler-88: Planetary parameters, \mstar, and \rstar are from this work, and other stellar parameters are from \citet{FultonPetigura2018}.  For 55 Cnc: stellar parameters are from \citet{VonBraun2011}, and planetary parameters are from \citet[][planet e]{Demory2016} and \citet[][other planets]{Baluev2015}.  For WASP-47: stellar and planetary parameters are from \citet{Vanderburg2017}.} 
\end{deluxetable*}

%
%

Like Kepler-88, WASP-47 has a nearly-circular hot Jupiter and a slightly eccentric longer-period giant planet \citep{Sinukoff2017_W47, Weiss2017, Vanderburg2017}.  Similarly, 55 Cnc has a close-in, nearly circular warm Jupiter at $P=14.7$ days, and three other known giant planets at 44.4, 261, and 4800 days \citep{Marcy2002,Naef2004,McArthur2004,Fischer2008,Wright2009,Dawson2010,Endl2012,Nelson2014,Baluev2015}.  In general, systems with hot Jupiters do not tend to have companions within $\sim1$ AU \citep{Steffen2012_HJ}, although many such systems have companions from 5-20 AU \citep{Bryan2016}.  Perhaps systems with hot/warm Jupiters in proximity to small exoplanets and/or with metal-rich stars are an exception to these patterns in the broader population. Long-baseline RV studies of the Kepler and TESS systems with hot and/or warm Jupiters in addition to small planets will reveal whether these systems also have distant giant planets.

Kepler-88 differs from 55 Cnc and WASP-47 in that the innermost known planet has an orbital period of 11 days, rather than $<1$ day.  Both 55 Cnc e and WASP-47 e are examples of ``ultra-short period'' (USP) planets, which are defined as having $P < 1$ day \citep{Dawson2010,Becker2015}.  USPs are generally small \citep[$\rpl < 2\,\rearth$][]{Sanchis-Ojeda2014}; this is likely because their high equilibrium temperatures do not support a volatile envelope of hydrogen and helium, although they could support thin envelopes of heavier mean molecular weight species like water or silicates \citep{Lopez2016}.  Kepler-88 b is nearly the size of Neptune, and, at a density of 1.1\,\gcc, has abundant hydrogen and helium.  However, its mass of \Mbval\ is very similar to the masses of 55 Cnc e and WASP-47 e.  We speculate that perhaps the nearby giant planet may eventually perturb Kepler-88 b into an orbit with $P < 1$ day, where photo-evaporation would remove the H/He envelope.  One mechanism that could accomplish this rearrangement is that the outer gas giant could perturb the 2:1 resonance to ever-widening libration amplitude, whereupon a scattering interaction between Kepler-88 b and c sends Kepler-88 b on an eccentric orbit which is tidally circularized at a much smaller orbital period. Kepler-88 is signficantly younger than the other systems (age = $1.8\pm1.6$ Gyr, see Table \ref{tab:star-compare}).  Thus, Kepler-88 might represent an early prototype of the 55 Cnc and WASP-47 systems, in particular their inclusion of a hot super-Earth.

\subsection{Comparison of RV and TTV Masses}

The measurement of planetary masses with radial velocity and transit timing
may always be confused by the presence of additional planets perturbing
the star (RV) or the transiting planet(s) (TTVs).  In addition, both techniques
can suffer from stellar systematics such as stellar jitter (RV) or stellar
photometric inhomogeneities (TTVs), as well as instrumental systematics.
This motivates a comparison of these two techniques when both are available.
Unfortunately the number of systems for which this is possible is very
in small in number  due to the small probability of transit 
of RV-detected systems, and the poor RV precision of most TTV systems found 
with Kepler.  \citet{Mills2017} found only nine planets which had both RV
and TTV mass measurements, of which eight agree to better than 2-$\sigma$, while
one (Kepler-89d) may be influenced by the presence of additional undiagnosed planets \citep{Mayo2017}.  
Note that each technique has a slightly different
dependence upon the stellar mass, so that precise stellar parameters are
needed to carry out a comparison.

The Kepler-88 system adds another planet for which both RV and TTV measurements
are available:  Kepler-88c.  In our RV analysis, we found a minimum mass of this planet
of $M_c \sin{i_c} = \McRV$, while in an analysis of the TTVs only using a three-planet model we found a best-fit mass of $M_c = 218\,\mearth$.  The photodynamical analysis indicates an
inclination for planet c of 93.14 degrees, so these two determinations agree
to within 1-$\sigma$.
Thus, Kepler-88c adds to the number of cases in which a consistent RV and TTV
mass are obtained, building confidence in both techniques.

\subsection{Opportunities for future observation}
The presence of two planets in near-resonant orbits can sometimes be confused for a single planet with moderate eccentricity \citep{Wittenmyer2019}.  Our RV baseline is too short, and our sampling of the periastron too sparse, to fully explore the possibility of a fourth planet in the system (see Figure \ref{fig:koi142-pgram}).  Our most recent RV just after the 2020 periastron passage of planet d shows no sign of a fourth planet.

The TESS spacecraft observed Kepler-88 during its northern hemisphere campaign this summer.  The photometric precision of TESS should be adequate to detect Kepler-88 b, if the planet is still transiting \citep{Christ2018}.

\section{Conclusion}
With six years of radial velocity monitoring, we have confirmed the presence, orbit, and mass of the giant planet Kepler-88 c: $P_c = \Pcval$, $M_c = \Mcval$.  This giant planet perturbs the orbit of the transiting planet Kepler-88 b and produces its TTVs.  We have also discovered an additional giant planet, Kepler-88 d, in an orbital period of $P_d = \Pdval$ with moderate eccentricity $e_d = \edval$ and mass $\msini_d = \mdval$.  Our analysis of the RVs only versus a full photodynamical model demonstrated that the RVs were necessary to detect planet d, but that the orbits and masses of planets b and c are much better determined with a full photodynamical model than with RVs alone.  Both techniques independently give consistent values for the mass of planet c. Kepler-88 joins the ranks of metal-rich stars that host both small transiting planets and two or more giant planets.

\acknowledgments
The authors thank the NASA Kepler Science team for providing the exquisite photometry on which this analysis is based.  Without the success of the Kepler Mission, the observational follow-up presented here would not have occurred.

LMW thanks Molly Kosiarek, Ian Crossfield, Sarah Blunt, Lee Rosenthal, Sam Grunblatt, and Ryan Rubenzahl for contributing to the collection of Keck-HIRES RVs, and Jason Wang, Dan Foreman-Mackey, and Ben Montet for useful discussions.  

LMW acknowledges the support of Ken \& Gloria Levy, the Trottier Family Foundation, the Beatrice Watson Parrent Fellowship, and NASA ADAP Grant 80NSSC19K0597.
DF acknowledges support from NASA grant NNX17AB93G.  EA acknowledges support from NSF grant AST-1615315, NASA grants NNX13AF62G, NNA13AA93A and 80NSSC18K0829.
Kepler was competitively selected as the tenth NASA Discovery mission. Funding for this mission is provided by the NASA Science Mission Directorate. 
We are grateful to the time assignment committees of the University of California, the University of Hawaii, and NASA for allocating observing time to complete this multi-year project. 
This work benefited from the 2017 Exoplanet Summer Program in the Other Worlds Laboratory (OWL) at the University of California, Santa Cruz, a program funded by the Heising-Simons Foundation. 
This research was partially conducted during the Exostar19 program at the Kavli Institute for Theoretical Physics at UC Santa Barbara, which was supported in part by the National Science Foundation under Grant No. NSF PHY-1748958.

The authors wish to recognize and acknowledge the very significant cultural role and
reverence that the summit of Mauna Kea has always had within the indigenous Hawaiian
community. We are most fortunate to have the opportunity to conduct observations from
this mountain.

\facilities{Kepler, Keck:I (HIRES)}
\software{RadVel (Fulton et al. 2018), TTVFast (Deck et al. 2014), Phodymm (Mills et al. 2016), Rebound (Rein \& Tamayo 2015)}

\clearpage
\bibliography{koi142_refs}{}
\bibliographystyle{aasjournal}

\end{document}